\documentclass[a4paper,fleqn]{cas-sc}

\usepackage[numbers,sort&compress]{natbib}
\usepackage{amsmath,amssymb}
\usepackage{array}
\usepackage{bm}
\usepackage{booktabs}
\usepackage{graphicx}
\usepackage{iftex}
\usepackage[section]{placeins}
\usepackage{xspace}

\ifPDFTeX
  \newcommand{\TableTimesFont}{\fontfamily{ptm}\selectfont}
\else
  \usepackage{fontspec}
  \newfontfamily\TableTimesFont{Times New Roman}
\fi

\AtBeginEnvironment{table}{\TableTimesFont}
\AtBeginEnvironment{tabular}{\TableTimesFont}

\ExplSyntaxOn
\cs_set:Npn \__make_tbl_caption:nn #1#2
{
  \l_tbl_align_tl
  \skip_vertical:N \l_tbl_abovecap_skip
  {\parbox{ \dimexpr(\l_tbl_width_dim)}
    {\rightskip=0pt\TableTimesFont\small\textbf{\color{scolor}#1}\par#2\par\vskip4pt }}
  \skip_vertical:N \l_tbl_belowcap_skip
}
\cs_set:Npn \__make_fig_caption:nn #1#2
{
  \l_fig_align_tl
  \skip_vertical:N \l_fig_abovecap_skip
  \setbox\cascaptionbox=\hbox{%
     \TableTimesFont\small\textbf{\color{scolor}#1:}~#2}
  \ifdim\the\wd\cascaptionbox<\dim_use:N \l_fig_width_dim\relax
    \parbox{ \l_fig_width_dim }
      {\unskip\ignorespaces\hfil\TableTimesFont\small
       \textbf{\color{scolor}#1:}~#2\hfil\par }
  \else
    \parbox{ \l_fig_width_dim }
      {\rightskip=0pt\unskip\ignorespaces\TableTimesFont
       \small\textbf{\color{scolor}#1:}~#2\par }
  \fi
  \skip_vertical:N \l_fig_belowcap_skip
}
\ExplSyntaxOff

\def\tsc#1{\csdef{#1}{\textsc{\lowercase{#1}}\xspace}}
\tsc{LBM}

\newcommand{\Grad}{\nabla_{\!X}}
\newcommand{\Div}{\nabla_{\!X}\!\cdot}
\newcommand{\tr}{\operatorname{tr}}
\newcommand{\Id}{\bm I}
\newcommand{\bX}{\bm X}
\newcommand{\bx}{\bm x}
\newcommand{\bu}{\bm u}
\newcommand{\bv}{\bm v}
\newcommand{\bb}{\bm b}
\newcommand{\bF}{\bm F}
\newcommand{\bG}{\bm G}
\newcommand{\bC}{\bm C}
\newcommand{\bE}{\bm E}

\newcommand{\bT}{\bm T}
\newcommand{\bP}{\bm P}
\newcommand{\bU}{\bm U}
\newcommand{\bB}{\bm B}
\newcommand{\bPhi}{\bm\Phi}
\newcommand{\bzero}{\bm 0}
\newcommand{\bn}{\bm n}
\newcommand{\bN}{\bm N}
\newcommand{\bH}{\bm H}
\newcommand{\dd}{\mathrm{d}}

\begin{document}
\let\WriteBookmarks\relax
\renewcommand{\topfraction}{0.95}
\renewcommand{\bottomfraction}{0.85}
\renewcommand{\textfraction}{0.05}
\renewcommand{\floatpagefraction}{0.80}
\setcounter{topnumber}{3}
\setcounter{bottomnumber}{2}
\setcounter{totalnumber}{5}
\setlength{\floatsep}{10pt plus 2pt minus 2pt}
\setlength{\textfloatsep}{12pt plus 2pt minus 2pt}
\setlength{\intextsep}{10pt plus 2pt minus 2pt}
\newcommand{\MainFigureWidth}{0.92\textwidth}
\newcommand{\StackedFigureWidth}{0.76\textwidth}

\shorttitle{Total-Lagrangian vectorial LBM for finite-strain hyperelastic dynamics}
\shortauthors{J. Feng and X. Chu}

\title[mode=title]{A total-Lagrangian vectorial lattice Boltzmann method for finite-strain hyperelastic dynamics}

\author[1]{Jingsen Feng}

\author[1]{Xu Chu}
\cormark[1]
\ead{x.chu@exeter.ac.uk}

\affiliation[1]{
  organization={Department of Engineering, University of Exeter},
  city={Exeter},
  postcode={EX4 4QF},
  country={United Kingdom}
}

\cortext[1]{Corresponding author: Xu Chu, x.chu@exeter.ac.uk}

\begin{abstract}
Inspired by the vectorial lattice Boltzmann method for linear elastodynamics \citep{boolakee2025linear}, we construct a total-Lagrangian vectorial lattice Boltzmann formulation for two-dimensional finite-strain hyperelastic dynamics. The governing equations are first written as a conservative first-order system for the material velocity and the full deformation gradient. This representation separates the kinematic part of the dynamics from the constitutive closure: the first Piola--Kirchhoff stress is evaluated locally from the current deformation gradient and enters the lattice only through nonlinear flux moments. A D2Q4 stencil with six-component vector populations is then used to match the state and the two material-coordinate fluxes. The formulation includes a second-order population initialization, trapezoidally centered body forcing, displacement reconstruction by velocity quadrature, and half-way reconstructions for velocity Dirichlet and Neumann traction boundaries on grid-aligned domains. The resulting method preserves the local collide--stream structure of standard lattice Boltzmann schemes while adapting the vectorial first-order strategy from linear elastodynamics to hyperelastic finite-strain dynamics.
\end{abstract}


\maketitle

\section{Introduction}
\label{sec:introduction}

The lattice Boltzmann method (LBM) \citep{mcnamara1988,succi2001,kruger2017} is designed to approximate solutions to a simplified kinetic equation by evolving a finite set of populations associated with discrete microscopic velocities. The gas-kinetic origin of these populations gives the method a local collide--stream structure \citep{haslam2008coupled,zhou2012gpu,verdier2020performance,zhang2021mpm}. This structure is attractive for explicit time integration and parallel implementation. In its native form, the LBM recovers the nearly incompressible Navier--Stokes equations through statistical moments of the populations and has become a mature tool in fluid dynamics and transport \citep{feng2023horizontal,liu2024homogenized,feng2026helmholtz,feng2026entropic}. These algorithmic features have also motivated attempts to use LBM in solid mechanics.

The extension of LBM to solids is motivated by the possibility of retaining locality, explicitness, and favorable parallel scaling for continuum-mechanics problems. A natural starting point is linear elasticity. For linear elastostatics, Yin et al. \citep{yin2016elastic} proposed an LBM based on displacement distribution functions for the linear elastic Lam\'e equation; this line of work was later characterized as a diffusion-type pseudo-time formulation whose steady state solves the elliptic equations of linear elastostatics \citep{boolakee2025linear}. Building on this idea, Boolakee et al. developed second-order accurate lattice Boltzmann schemes for quasi-static linear elasticity and corresponding second-order Dirichlet and Neumann boundary formulations on arbitrary curved two-dimensional domains \citep{boolakee2023static,boolakee2023bc}. These works demonstrated that LBM can be constructed systematically for linear elastostatic boundary-value problems. At the same time, as noted in subsequent work \citep{boolakee2025linear}, the need to march in pseudo-time until a steady state is reached limits the competitiveness of such approaches for static problems when compared with established finite element methods.

For solid dynamics, especially problems involving elastic-wave propagation, the algorithmic structure of LBM is more naturally aligned with the target equations. Early attempts include the lattice Boltzmann model for solid-body dynamics of Marconi and Chopard \citep{marconi2003solid} and the elastic-wave formulation of O'Brien et al. for Poisson solids \citep{obrien2012poisson}. Murthy et al. \citep{murthy2018elasticwaves} later introduced a lattice Boltzmann formulation for elastic wave propagation with a tunable Poisson ratio, based on a moment-chain construction. Escande et al. \citep{escande2020elasticwaves} provided a theoretical and stability analysis of a regular-lattice version of this approach and validated it for bulk and surface waves, while related developments addressed crack loading and boundary conditions for elastodynamics \citep{schluter2018crack,faust2024bc}. These contributions showed that LBM can reproduce transient elastic responses and wave propagation in solids. Nevertheless, formulations based on scalar populations and standard velocity sets have limited freedom to represent the coupled flux structure of elasticity. In practice, moment-chain schemes have relied on additional moment equations, finite-difference corrections, or artificial dissipation to supply missing quantities and improve stability \citep{boolakee2025linear}. As a consequence, second-order consistency, stability for arbitrary material parameters, and accurate boundary treatment remain delicate issues \citep{faust2024bc,boolakee2025linear}.

Within this scalar-population moment-chain framework, a recent development has extended the approach to geometrically and constitutively nonlinear elastodynamics \citep{muller2025nonlinear}. In that formulation, stresses and deformation measures are evaluated in the reference configuration, and nonlinear constitutive behavior is incorporated through a forcing term. The method demonstrates the feasibility of LBM for finite-strain solid dynamics through benchmark problems such as uniaxial tension, simple shear, and bending waves. However, it still relies on finite-difference evaluations of gradients and divergences, and the nonlinear Piola stress does not enter the lattice through a direct moment representation of the physical fluxes. This leaves open the question of whether finite-strain hyperelastic dynamics can be formulated in a way that remains closer to the basic lattice Boltzmann algorithm while treating the solid-mechanics fluxes as primary moment-matched quantities.

In this paper, we pursue this direction using vectorial LBM. The key modification of a vectorial formulation is the use of vector-valued populations instead of scalar ones, while preserving the local collision and streaming steps. Vectorial kinetic formulations have been used in several contexts, including magnetohydrodynamics \citep{dellar2009}, hyperbolic conservation laws and relaxation schemes \citep{jin1995,graille2014,dubois2014vectorial}, and incompressible-flow discretizations with boundary and stability analysis \citep{zhao2024}. Most relevant to the present work, Boolakee et al. \citep{boolakee2025linear} recently introduced a vectorial LBM for linear elastodynamics by first rewriting the displacement equation as an equivalent first-order hyperbolic system. They established second-order consistency, stability estimates under a CFL-like condition, second-order population initialization, and Dirichlet boundary conditions on rectangular domains. This result indicates that vector-valued populations provide the additional degrees of freedom needed to represent the state and fluxes of coupled solid-dynamics systems.

We construct here a total-Lagrangian vectorial lattice Boltzmann formulation for two-dimensional finite-strain hyperelastic dynamics. The governing equations are first written as a conservative first-order system for the material velocity and the full deformation gradient. In this representation the lattice remains fixed in the reference configuration, while the first Piola--Kirchhoff stress is evaluated locally from the current deformation gradient and enters the method through nonlinear flux moments. A D2Q4 stencil with six-component vector populations is then used to match the macroscopic state and the two material-coordinate fluxes. The formulation includes trapezoidally centered body forcing, a second-order population initialization, displacement reconstruction by velocity quadrature, and half-way reconstructions for velocity Dirichlet and nominal-traction Neumann boundaries on grid-aligned domains. Numerical experiments verify the construction for manufactured solutions, finite-strain benchmark tests, several hyperelastic constitutive laws, acoustic-tensor wave speeds, finite-amplitude periodic waves, and a bounded-domain cantilever bending wave.

The manuscript is organized as follows. Section~\ref{sec:governing-equations} summarizes the total-Lagrangian finite-strain equations, hyperelastic closure, and first-order conservative form. Section~\ref{sec:d2q4-vectorial-lbm} introduces the D2Q4\(\times\)6 vectorial lattice Boltzmann discretization, including equilibrium moments, forcing, initialization, and boundary reconstruction. Section~\ref{sec:numerical-validation} presents the manufactured-solution and benchmark validation, while Section~\ref{sec:constitutive-wave-dynamics} examines constitutive response and finite-strain wave dynamics. Section~\ref{sec:conclusions-outlook} concludes with the main limitations and possible extensions.

\section{Total-Lagrangian finite-strain elastodynamics}
\label{sec:governing-equations}

This section prepares the total-Lagrangian equations for the vectorial lattice Boltzmann discretization. We work on a fixed reference domain \(\Omega_0\subset\mathbb R^2\) with material coordinate \(\bX=(X_1,X_2)^T=(X,Y)^T\). The current position is \(\bx(\bX,t)\), the displacement is \(\bu(\bX,t)=\bx(\bX,t)-\bX\), and the material velocity is \(\bv=\partial_t\bu\). We write \(\partial_A=\partial/\partial X_A\), use \(\Grad\bu\) for the material displacement gradient, and denote the identity tensor by \(\Id\). Latin indices \(i,k=1,2\) denote spatial components, capital indices \(A,B=1,2\) denote material-coordinate directions, and repeated indices are summed.

Let \(W(\bF)\) be the strain-energy density per unit reference volume, \(\bP(\bF)=\partial W/\partial\bF\) the first Piola--Kirchhoff stress, and \(\bb=(b_1,b_2)^T\) the prescribed body acceleration. In the nondimensional displacement variables used below, the hyperelastic equations can be written as \citep{ogden1997,holzapfel2000,bonetwood2008}
\begin{equation}
  \partial_t^2 u_i=\partial_A P_{iA}(\bF)+b_i,
  \qquad
  \bF=\Id+\Grad\bu,
  \qquad
  P_{iA}(\bF)=\frac{\partial W}{\partial F_{iA}}(\bF).
  \label{eq:displacement-form}
\end{equation}
Equation~\eqref{eq:displacement-form} combines material momentum balance with the hyperelastic constitutive map. The stress divergence depends on displacement gradients through \(\bF=\Id+\Grad\bu\), and the equation is second order in time. Classical LBM is naturally suited to first-order conservation and balance systems, since its local collision and streaming update is closed through moments representing macroscopic states and fluxes. For hyperelasticity, the corresponding step is to lower the order of Eq.~\eqref{eq:displacement-form}: we introduce \(\bv=\partial_t\bu\), evolve \(\bF\) directly, and obtain a first-order system. The vectorial LBM used below then carries this first-order system with vector-valued populations whose moments recover the state and the two material-coordinate fluxes.

\subsection{Scaling and reference description}
\label{subsec:nondim}

The total-Lagrangian description is the natural setting for a fixed Cartesian lattice: lattice nodes remain stationary, material points are indexed by their reference coordinates, and the stress divergence is written in terms of the first Piola--Kirchhoff stress. We nondimensionalize by a reference length \(L_0\), density \(\rho_0\), and velocity \(V_0\). Thus \(\bX=\bX^{\rm dim}/L_0\), \(\bx=\bx^{\rm dim}/L_0\), \(\bu=\bu^{\rm dim}/L_0\), \(t=V_0t^{\rm dim}/L_0\), and \(\bv=\bv^{\rm dim}/V_0\). The stress, body acceleration, and Lame parameters are scaled by
\[
  \bP=\frac{\bP^{\rm dim}}{\rho_0V_0^2},\qquad
  \bb=\frac{L_0\bb^{\rm dim}}{V_0^2},\qquad
  \lambda=\frac{\lambda^{\rm dim}}{\rho_0V_0^2},\qquad
  \mu=\frac{\mu^{\rm dim}}{\rho_0V_0^2}.
\]
The deformation gradient \(\bF=\partial\bx/\partial\bX\) is dimensionless because it is a ratio of two lengths. If \(V_0\) is chosen as the infinitesimal longitudinal wave speed \(c_{L,0}^{\rm dim}=\sqrt{(\lambda^{\rm dim}+2\mu^{\rm dim})/\rho_0}\), then \(\lambda+2\mu=1\). Other choices of \(V_0\) simply rescale the nondimensional moduli.

The lattice spacing and time step used in Section~\ref{sec:d2q4-vectorial-lbm} are \(\Delta x=\Delta X^{\rm dim}/L_0\) and \(\Delta t=V_0\Delta t^{\rm dim}/L_0\), so the nondimensional lattice speed is \(c=\Delta x/\Delta t=\Delta X^{\rm dim}/(V_0\Delta t^{\rm dim})\). All variables below are nondimensional.

\subsection{Kinematics, balance laws, and boundary conditions}
\label{subsec:kinematics-balance}

Finite-strain dynamics are carried by the deformation gradient \(\bF=\Id+\Grad\bu\) and its determinant \(J=\det\bF\). The admissible range is \(J>0\), since loss of orientation makes the hyperelastic stress and the Piola transformation ill-defined. In the present two-dimensional formulation \(\bF\in\mathbb R^{2\times2}\). Plane-strain or plane-stress reductions of a three-dimensional material require separate constitutive reductions, which are outside the scope of this formulation.

Introducing \(\bv=\partial_t\bu\) in Eq.~\eqref{eq:displacement-form} and evolving \(\bF=\Id+\Grad\bu\) directly gives the nondimensional first-order system
\begin{equation}
  \partial_t v_i=\partial_A P_{iA}(\bF)+b_i,
  \qquad
  \partial_tF_{iA}=\partial_Av_i.
  \label{eq:momentum-and-F-evolution}
\end{equation}
The first equation is the momentum balance from Eq.~\eqref{eq:displacement-form} written for velocity. The second is the kinematic identity obtained by differentiating \(F_{iA}=\delta_{iA}+\partial_Au_i\) in time. Thus Eq.~\eqref{eq:momentum-and-F-evolution} is the order-reduced form of the displacement equation. Once \(\bF\) is known at a lattice node, a hyperelastic stress can be evaluated locally; no finite-difference reconstruction of strain from displacement is needed in the bulk.

The initial conditions are prescribed as \(\bu(\bX,0)=\bu_0(\bX)\) and \(\bv(\bX,0)=\bv_0(\bX)\), which imply \(\bF(\bX,0)=\Id+\Grad\bu_0(\bX)\). This initialization automatically satisfies the compatibility relation \(\partial_XF_{i2}-\partial_YF_{i1}=0\) at \(t=0\). For smooth solutions, Eq.~\eqref{eq:momentum-and-F-evolution} preserves this compatibility at the continuum level.

Let the boundary be split into a displacement part \(\Gamma_D\) and a traction part \(\Gamma_N\). On \(\Gamma_D\), the physical condition is \(\bu=\bu_D\). The first-order state evolves velocity, so the lattice boundary condition uses the corresponding boundary velocity \(\bv_D=\partial_t\bu_D\). On \(\Gamma_N\), the Neumann condition is written in total-Lagrangian form as
\begin{equation}
  \bP(\bF)\bN=\bar{\bT}
  \qquad\text{on }\Gamma_N,
  \label{eq:neumann-continuum}
\end{equation}
where \(\bN\) is the outward unit normal to the reference boundary and \(\bar{\bT}\) is the prescribed nominal traction per unit reference length/area. This is the Neumann condition expressed in reference variables. The spatial traction \(\bm\sigma\bn=\bar{\bm t}\) and the nominal traction are related by Nanson's formula, \(\bm\sigma\bn\,\dd a=\bP\bN\,\dd A_0\) with \(\bP=J\bm\sigma\bF^{-T}\) \citep{holzapfel2000,bonetwood2008}. Thus the total-Lagrangian prescribed traction is \(\bP\bN\); a spatial traction \(\bm\sigma\bn\) must first be converted to the reference configuration.

The boundary reconstructions in Section~\ref{subsec:boundary-reconstruction} are written for grid-aligned rectangular boundaries, so \(\bN\in\{(1,0),(-1,0),(0,1),(0,-1)\}\). The half-way D2Q4 boundary construction imposes this algorithmic restriction.

\subsection{Hyperelastic closure}
\label{subsec:hyperelastic-models}

The lattice method requires only a pointwise map from deformation gradient to stress and, for initialization, wave-speed estimates, and traction-boundary Newton iterations, its tangent. For a hyperelastic material with strain-energy density \(W(\bF)\), these quantities are
\begin{equation}
  \bP(\bF)=\frac{\partial W}{\partial\bF},
  \qquad
  \mathbb C_{iAkB}(\bF)=\frac{\partial P_{iA}}{\partial F_{kB}},
  \qquad
  \delta P_{iA}=\mathbb C_{iAkB}\,\delta F_{kB}.
  \label{eq:stress-and-tangent}
\end{equation}
These two quantities play different roles in the first-order system below. The stress \(\bP(\bF)\) supplies the stress components of the material-coordinate flux vectors \(\bPhi_X\) and \(\bPhi_Y\) in Eq.~\eqref{eq:flux-vectors}. The tangent \(\mathbb C\) supplies the corresponding flux derivatives with respect to \(\bF\), which enter the Jacobian actions in Eq.~\eqref{eq:jacobian-actions}.

The numerical method does not depend on a particular strain-energy form. The implementation used in the examples includes St. Venant--Kirchhoff, compressible neo-Hookean, logarithmic neo-Hookean, Mooney--Rivlin, Yeoh, and Gent closures. Their explicit \(W(\bF)\) and \(\bP(\bF)\) are collected in Appendix~\ref{app:constitutive-laws}. In the main algorithm, any admissible hyperelastic law can be used as long as \(\bP(\bF)\) and the tangent actions in Eq.~\eqref{eq:stress-and-tangent} are available for \(J>0\). The local wave-speed estimates used below also assume that the tangent remains strongly elliptic over the simulated states.

\subsection{First-order system used by the lattice method}
\label{subsec:first-order-system}

The vectorial LBM will approximate Eq.~\eqref{eq:momentum-and-F-evolution} through moments of six-component populations. The macroscopic state is
\begin{equation}
  \bU=(v_1,v_2,F_{11},F_{12},F_{21},F_{22})^T.
  \label{eq:state-vector}
\end{equation}
With this ordering, the coupled velocity--deformation-gradient equations become
\begin{equation}
  \partial_t\bU+\partial_X\bPhi_X(\bU)+\partial_Y\bPhi_Y(\bU)=\bB,
  \qquad
  \bB=(b_1,b_2,0,0,0,0)^T,
  \label{eq:first-order-conservation-law}
\end{equation}
where the two material-coordinate fluxes are
\begin{equation}
  \bPhi_X(\bU)=(-P_{11},-P_{21},-v_1,0,-v_2,0)^T,
  \qquad
  \bPhi_Y(\bU)=(-P_{12},-P_{22},0,-v_1,0,-v_2)^T.
  \label{eq:flux-vectors}
\end{equation}
The signs follow from writing the equations in the conservative form \(\partial_t\bU+\partial_A\bPhi_A=\bB\). Taking the divergence of \(\bPhi_X\) and \(\bPhi_Y\) gives \(-\partial_AP_{iA}\) in the velocity rows and \(-\partial_Av_i\) in the deformation-gradient rows, so Eq.~\eqref{eq:first-order-conservation-law} is exactly Eq.~\eqref{eq:momentum-and-F-evolution}.

This representation is the finite-strain analogue of the first-order linear elastodynamic system used in vectorial LBM. The essential change is that the stress entries in \(\bPhi_X\) and \(\bPhi_Y\) are nonlinear functions of the last four components of \(\bU\). Consequently, the flux Jacobians \(\bm A_X=\partial\bPhi_X/\partial\bU\) and \(\bm A_Y=\partial\bPhi_Y/\partial\bU\) are state dependent. Their action on an increment \(\delta\bU=(\delta v_1,\delta v_2,\delta F_{11},\delta F_{12},\delta F_{21},\delta F_{22})^T\) is
\begin{equation}
  \bm A_X\delta\bU=(-\delta P_{11},-\delta P_{21},-\delta v_1,0,-\delta v_2,0)^T,
  \quad
  \bm A_Y\delta\bU=(-\delta P_{12},-\delta P_{22},0,-\delta v_1,0,-\delta v_2)^T,
  \label{eq:jacobian-actions}
\end{equation}
with \(\delta P_{iA}=\mathbb C_{iAkB}\delta F_{kB}\). These Jacobian--vector products are sufficient for initialization and local wave-speed estimates, allowing the implementation to avoid dense \(6\times6\) matrices.

For a unit material direction \(\bn\), define the total-Lagrangian acoustic tensor \citep{ogden1997,holzapfel2000}
\begin{equation}
  Q_{ik}(\bn;\bF)=n_A\mathbb C_{iAkB}(\bF)n_B.
  \label{eq:acoustic-tensor-section2}
\end{equation}
When the material is strongly elliptic, \(\bm a\cdot Q(\bn;\bF)\bm a>0\) for all nonzero \(\bm a\) and all unit \(\bn\). The eigenvalues of \(Q\) are the squared local tangent wave speeds in direction \(\bn\), after the reference density has been absorbed by nondimensionalization. This quantity is used in Section~\ref{sec:numerical-validation} to audit the lattice speed against the largest tangent wave speed encountered in the validation cases.

\section{D2Q4\texorpdfstring{$\times$}{x}6 vectorial lattice Boltzmann discretization}
\label{sec:d2q4-vectorial-lbm}

The previous section reduced finite-strain hyperelastic dynamics to a first-order system with one state vector and two physical fluxes. The vectorial LBM now has a clear task: store enough population information at each node to reconstruct \(\bU\), \(\bPhi_X(\bU)\), and \(\bPhi_Y(\bU)\) by moments. This is the same design principle used in the linear vectorial elastodynamic scheme, where vector-valued populations enable a first-order hyperbolic system to be approximated while preserving the local collide--stream structure of LBM \citep{boolakee2025linear}. The finite-strain extension keeps the algebraic moment structure and replaces constant elastic fluxes by the nonlinear Piola fluxes.

\subsection{Lattice populations and what their moments represent}
\label{subsec:d2q4-velocity-set}

The material domain is discretized by a uniform Cartesian lattice with spacing \(\Delta x=\Delta y\) and time step \(\Delta t\). The four D2Q4 directions are \(\mathcal V=\{(1,0),(0,1),(-1,0),(0,-1)\}\), and the velocity associated with \(q=(i,j)\) is \(\bm c_q=c(i,j)\), where \(c=\Delta x/\Delta t\). For each direction, the method stores a six-component population \(\bm f_{ij}(\bX,t)\in\mathbb R^6\). Here ``population'' denotes a numerical carrier of moments, with no gas-kinetic molecular interpretation for the solid model.

Each D2Q4 population is a six-component vector, so the four lattice directions provide four independent vector combinations. The continuum system requires three of them: the zeroth moment gives the state, and the two first moments give the \(X\)- and \(Y\)-fluxes. The remaining vector degree of freedom has no continuum counterpart; we fix it by setting the symmetric difference between the horizontal and vertical population pairs to zero. This gives a unique nearest-neighbour equilibrium. With body acceleration, the state is recovered as
\begin{equation}
  \bU^{\rm num}(\bX,t)=\sum_{(i,j)\in\mathcal V}\bm f_{ij}(\bX,t)+\frac{\Delta t}{2}\bB(\bX,t).
  \label{eq:moment-U}
\end{equation}
The half-step source shift is the discrete analogue of trapezoidal source centering. It makes the body force contribute to the recovered state without contaminating the flux moments.

\subsection{Equilibrium distribution by moment matching}
\label{subsec:equilibrium-moment-constraints}

The equilibrium populations are defined by requiring their moments to reproduce the continuum quantities:
\begin{equation}
  \sum_q\bm f_q^{eq}=\bU,
  \qquad
  \sum_q c i\,\bm f_q^{eq}=\bPhi_X(\bU),
  \qquad
  \sum_q c j\,\bm f_q^{eq}=\bPhi_Y(\bU),
  \qquad
  \sum_q c^2(i^2-j^2)\bm f_q^{eq}=\bzero.
  \label{eq:equilibrium-moments}
\end{equation}
Here and below \(q=(i,j)\), and \(\sum_q\) means summation over \(\mathcal V\). Solving these four vector equations gives the compact expression
\begin{equation}
  \bm f_{ij}^{eq}(\bU)=\frac14\left[\bU+\frac{2}{c}\{i\bPhi_X(\bU)+j\bPhi_Y(\bU)\}\right],
  \qquad (i,j)\in\mathcal V.
  \label{eq:feq}
\end{equation}
This equation is the central algebraic object of the method. It says that the equilibrium distribution is the lowest-order D2Q4 interpolation of one state and two fluxes. The finite-strain mechanics enter only through the evaluation of \(\bPhi_X\) and \(\bPhi_Y\), i.e. through \(\bP(\bF)\).

Because the fluxes are nonlinear, the derivative of the equilibrium is needed whenever the method is expanded or initialized to second order. For any increment \(\bm Z\in\mathbb R^6\),
\begin{equation}
  (\bm f_{ij}^{eq})'(\bU)\bm Z=\frac14\left[\bm Z+\frac{2}{c}\{i\bm A_X(\bU)\bm Z+j\bm A_Y(\bU)\bm Z\}\right].
  \label{eq:feq-derivative}
\end{equation}
Taking moments of this identity recovers \(\bm Z\), \(\bm A_X\bm Z\), and \(\bm A_Y\bm Z\). Thus the equilibrium derivative carries exactly the linearized continuum flux information needed for a Taylor expansion of the lattice update.

\subsection{Collision, streaming, and source centering}
\label{subsec:collision-streaming-forcing}

A time step begins by evaluating \(\bU^{\rm num}\) from Eq.~\eqref{eq:moment-U}. The last four entries define \(\bF\); the selected hyperelastic model then gives \(\bP(\bF)\), and hence the fluxes in Eq.~\eqref{eq:flux-vectors}. The BGK collision step \citep{bhatnagar1954,succi2001,kruger2017} relaxes each population toward the local equilibrium:
\begin{equation}
  \bm f_{ij}^{*}=\bm f_{ij}+\omega(\bm f_{ij}^{eq}(\bU^{\rm num})-\bm f_{ij})+\Delta t(2-\omega)w_{ij}\bB,
  \qquad
  w_{ij}=\frac18.
  \label{eq:collision}
\end{equation}
The post-collision population is then streamed to the neighbouring node in direction \((i,j)\):
\begin{equation}
  \bm f_{ij}\bigl(\bX+c(i,j)\Delta t,t+\Delta t\bigr)=\bm f_{ij}^{*}(\bX,t).
  \label{eq:streaming}
\end{equation}
The symmetric weights satisfy \(\sum_qw_q=1/2\), \(\sum_qciw_q=0\), and \(\sum_qcjw_q=0\). Thus the source enters the zeroth moment consistently with Eq.~\eqref{eq:moment-U}, leaving the stress and kinematic flux moments unchanged.

The nondissipative second-order setting is \(\omega=2\), for which the explicit forcing term in Eq.~\eqref{eq:collision} vanishes because the source has already been time-centred in the state recovery. Values \(1<\omega<2\) introduce kinetic dissipation, which can be useful in strongly nonlinear computations. The formal second-order nondissipative expansion is tied to \(\omega=2\); the consistency argument is collected in Appendix~\ref{app:second-order-consistency}.

\subsection{Second-order population initialization}
\label{subsec:second-order-initialization}

The prescribed initial fields determine \(\bU_0\) and leave the four populations underdetermined. An equilibrium initialization, \(\bm f_q(\bX,0)=\bm f_q^{eq}(\bU_0)\), recovers the leading macroscopic state. The non-equilibrium part expected by the lattice dynamics is then absent, and the resulting initial kinetic layer reduces short-time accuracy. For smooth initial fields, following the asymptotic initialization used in vectorial LBM \citep{boolakee2025linear,junk2005}, the leading correction follows by expanding the distribution one half time step backward along its lattice characteristic:
\begin{equation}
  \bm f_q(\bX,0)=\bm f_q^{eq}(\bU_0)-\frac{\Delta t}{2}D_q\bm f_q^{eq}(\bU_0),
  \qquad
  D_q=\partial_t+ci\partial_X+cj\partial_Y.
  \label{eq:init-compact}
\end{equation}
No additional time derivative is prescribed. The continuum equation gives \(\partial_t\bU\), so the derivative in Eq.~\eqref{eq:init-compact} can be evaluated from spatial derivatives of the initial fields. For \(q=(i,j)\), set
\begin{equation}
  \bm Z_q=D_q\bU_0
  =\bigl(Z_1,Z_2,Z_3,Z_4,Z_5,Z_6\bigr)^T,
  \label{eq:init-Zq}
\end{equation}
where all quantities on the right are evaluated at \(t=0\), and
\begin{align}
  Z_1&=b_1+\partial_XP_{11}+\partial_YP_{12}+ci\,\partial_Xv_1+cj\,\partial_Yv_1, \notag\\
  Z_2&=b_2+\partial_XP_{21}+\partial_YP_{22}+ci\,\partial_Xv_2+cj\,\partial_Yv_2, \notag\\
  Z_3&=\partial_Xv_1+ci\,\partial_XF_{11}+cj\,\partial_YF_{11}, \notag\\
  Z_4&=\partial_Yv_1+ci\,\partial_XF_{12}+cj\,\partial_YF_{12}, \notag\\
  Z_5&=\partial_Xv_2+ci\,\partial_XF_{21}+cj\,\partial_YF_{21}, \notag\\
  Z_6&=\partial_Yv_2+ci\,\partial_XF_{22}+cj\,\partial_YF_{22}.
  \label{eq:init-Zq-components}
\end{align}
The stress derivatives in Eq.~\eqref{eq:init-Zq-components} are computed from the material tangent, for example \(\partial_\alpha P_{iA}=\mathbb C_{iAkB}\partial_\alpha F_{kB}\) for \(\alpha=X,Y\). Define
\begin{equation}
  \bm Z_F=
  \begin{pmatrix}
    Z_3 & Z_4\\
    Z_5 & Z_6
  \end{pmatrix}.
  \label{eq:init-ZF}
\end{equation}
The stress increment induced by \(\bm Z_F\) is \(\delta P^q_{iA}=\mathbb C_{iAkB}Z_{F,kB}\), or componentwise
\begin{align}
  \delta P^q_{11}&=\mathbb C_{1111}Z_3+\mathbb C_{1112}Z_4+\mathbb C_{1121}Z_5+\mathbb C_{1122}Z_6, \notag\\
  \delta P^q_{12}&=\mathbb C_{1211}Z_3+\mathbb C_{1212}Z_4+\mathbb C_{1221}Z_5+\mathbb C_{1222}Z_6, \notag\\
  \delta P^q_{21}&=\mathbb C_{2111}Z_3+\mathbb C_{2112}Z_4+\mathbb C_{2121}Z_5+\mathbb C_{2122}Z_6, \notag\\
  \delta P^q_{22}&=\mathbb C_{2211}Z_3+\mathbb C_{2212}Z_4+\mathbb C_{2221}Z_5+\mathbb C_{2222}Z_6.
  \label{eq:init-deltaPq}
\end{align}
Using Eq.~\eqref{eq:feq-derivative}, the correction is then the explicit vector
\begin{equation}
  D_q\bm f_q^{eq}(\bU_0)
  =\frac14
  \begin{pmatrix}
    Z_1-\frac{2}{c}\bigl(i\,\delta P^q_{11}+j\,\delta P^q_{12}\bigr)\\
    Z_2-\frac{2}{c}\bigl(i\,\delta P^q_{21}+j\,\delta P^q_{22}\bigr)\\
    Z_3-\frac{2i}{c}Z_1\\
    Z_4-\frac{2j}{c}Z_1\\
    Z_5-\frac{2i}{c}Z_2\\
    Z_6-\frac{2j}{c}Z_2
  \end{pmatrix}.
  \label{eq:init-expanded}
\end{equation}
For manufactured solutions the derivatives are obtained analytically. For benchmark initial conditions, the same formula is evaluated using second-order finite differences, with one-sided stencils next to nonperiodic boundaries.

\subsection{Displacement and stress recovery}
\label{subsec:post-processing}

The lattice evolves velocity and deformation gradient. Displacement is reconstructed only for output and for displacement error norms. A trapezoidal update is used: with \(\bu^{*,0}=\bu_0-(\Delta t/2)\bv_0\), set
\begin{equation}
  \bu^n=\bu^{*,n}+\frac{\Delta t}{2}\bv^n,
  \qquad
  \bu^{*,n+1}=\bu^n+\frac{\Delta t}{2}\bv^n.
  \label{eq:trapezoid-u}
\end{equation}
The stress used by the dynamics is always the first Piola stress \(\bP(\bF)\). When Cauchy stress is needed for diagnostics or comparison with spatial benchmark fields, it is recovered by \(\bm\sigma=J^{-1}\bP\bF^T\).

\subsection{Half-way boundary reconstruction}
\label{subsec:boundary-reconstruction}

The boundary treatment should be read from the viewpoint of the lattice. At a boundary node, the macroscopic equation remains available; after streaming, one or more incoming populations are unknown. The role of the boundary condition is therefore to reconstruct each missing incoming population from the known outgoing population and the imposed physical boundary condition.

For a grid-aligned boundary, let \(\bN\) be the outward reference normal. The missing incoming direction is \(d=(i,j)=-\bN\), and \(-d=\bN\) is the opposite outgoing direction. Following link-wise boundary reconstructions used in elastic-solid LBM \citep{boolakee2023bc,faust2024bc,boolakee2025linear}, we use a half-way reconstruction of the form
\begin{equation}
  \bm f_d(t+\Delta t)=\bm D\bm f_{-d}^{*}(t)+\bm S_d(t+\Delta t/2).
  \label{eq:boundary-general}
\end{equation}
The signs in \(\bm D\) are fixed by the pair identities obtained from Eq.~\eqref{eq:feq}: \(\bm f_d^{eq}+\bm f_{-d}^{eq}=\bU/2\) and \(\bm f_d^{eq}-\bm f_{-d}^{eq}=d_A\bPhi_A(\bU)/c\). Thus anti-bounce-back imposes a state component, and bounce-back with a correction imposes a normal flux component.

\subsubsection{Velocity Dirichlet boundary condition}
\label{subsubsec:velocity-dirichlet-boundary}

For a prescribed displacement \(\bu_D\), the first-order state sees \(\bv_D=\partial_t\bu_D\). Anti-bounce-back reconstructs the velocity entries of the missing population. Bounce-back reconstructs the deformation-gradient entries with the compatible kinematic flux inserted. The rule is
\begin{equation}
  \bm D_D=\operatorname{diag}(-1,-1,1,1,1,1),
  \qquad
  \bm S_d^D=\left(\frac{v_{D1}}{2},\frac{v_{D2}}{2},\frac{N_1v_{D1}}{c},\frac{N_2v_{D1}}{c},\frac{N_1v_{D2}}{c},\frac{N_2v_{D2}}{c}\right)^T.
  \label{eq:dirichlet-boundary-rule}
\end{equation}
The first two entries impose the boundary velocity. The last four entries express the normal kinematic fluxes of \(F_{11},F_{12},F_{21},F_{22}\). Hence a displacement boundary enters through \(\partial_t\bu_D\), because the evolved lattice variables are velocity and deformation gradient.

\subsubsection{Neumann boundary condition}
\label{subsubsec:neumann-boundary}

For a total-Lagrangian Neumann boundary, the prescribed datum is the nominal traction \(\bar{\bT}=\bP\bN\). Since \(d=-\bN\), the normal flux of the velocity rows satisfies \(d_A\bPhi_A=(\bP\bN,0,0,0,0)^T\) in its first two components. Bounce-back plus the traction correction reconstructs the velocity entries. Anti-bounce-back reconstructs the deformation-gradient entries using a boundary deformation gradient \(\bF^b\):
\begin{equation}
  \bm D_N=\operatorname{diag}(1,1,-1,-1,-1,-1),
  \qquad
  \bm S_d^N=\left(\frac{\bar T_1}{c},\frac{\bar T_2}{c},\frac{F_{11}^b}{2},\frac{F_{12}^b}{2},\frac{F_{21}^b}{2},\frac{F_{22}^b}{2}\right)^T.
  \label{eq:neumann-boundary-rule}
\end{equation}
The unresolved part is \(\bF^b\). At a boundary whose normal is aligned with material direction \(A\), the tangential column of \(\bF^b\) is extrapolated from interior nodes; the normal column is determined from the two scalar traction equations \(P_{iA}(\bF^b)N_A=\bar T_i\), \(i=1,2\). We solve this local nonlinear system by Newton iteration with line search and reject trial states with \(\det\bF^b\le0\). The procedure introduces no global boundary solve; the nonlinearity is confined to a two-unknown constitutive inversion at each boundary link.

The traction \(\bar{\bT}\) is measured per reference area. Loads prescribed per current area, such as a spatial traction \(\bm\sigma\bn\), are pulled back to the reference boundary before they enter Eq.~\eqref{eq:neumann-boundary-rule}.

\subsection{One step of the method}
\label{subsec:one-time-step}

At the \(n\)-th time step, the update consists of the following local operations:
\begin{enumerate}
  \item Recover the macroscopic state \(\bU^{{\rm num},n}=\sum_q\bm f_q^n+(\Delta t/2)\bB^n\). Its first two components give \(\bv^n\), and the last four define \(\bF^n=(F_{11}^n,F_{12}^n,F_{21}^n,F_{22}^n)\).
  \item Evaluate the constitutive response \(\bP^n=\bP(\bF^n)\), then assemble the material-coordinate fluxes \(\bPhi_X^n=\bPhi_X(\bU^{{\rm num},n})\) and \(\bPhi_Y^n=\bPhi_Y(\bU^{{\rm num},n})\).
  \item Form the equilibrium populations \(\bm f_q^{{\rm eq},n}=\bm f_q^{eq}(\bU^{{\rm num},n})\) and apply the collision rule to obtain \(\bm f_q^{*,n}\).
  \item Stream each post-collision population along \(\bm c_q=c(i,j)\), i.e. \(\bm f_q^{n+1}(\bX+\bm c_q\Delta t)=\bm f_q^{*,n}(\bX)\).
  \item Complete boundary nodes by reconstructing each missing incoming population, \(\bm f_d^{n+1}=\bm D\bm f_{-d}^{*,n}+\bm S_d^{n+1/2}\), using the periodic, Dirichlet, or Neumann rule.
  \item Update output variables with \(\bu^n=\bu^{*,n}+(\Delta t/2)\bv^n\) and \(\bu^{*,n+1}=\bu^n+(\Delta t/2)\bv^n\); when needed, recover \(\bm\sigma^n=(J^n)^{-1}\bP^n(\bF^n)^T\).
\end{enumerate}

The first moments of the equilibrium populations are constrained to equal \(\bPhi_X\) and \(\bPhi_Y\), so the streaming step applies the lattice stencil directly to these fluxes. In the velocity rows this gives the stress-divergence term \(\partial_AP_{iA}\).

\section{Accuracy against reference solutions and finite-strain benchmarks}
\label{sec:numerical-validation}

This section assesses the formulation against reference solutions with increasing numerical complexity. The first two tests use exact manufactured fields generated by the method of manufactured solutions (MMS) and isolate the bulk nonlinear fluxes, the population initialization, and the half-way boundary reconstructions. The last two tests reproduce the uniaxial-tension and simple-shear benchmarks of M\"uller et al.~\citep{muller2025nonlinear} using an independent finite-element reference on the same benchmark definitions. The sequence is intended to build the numerical evidence from controlled reference fields to established finite-strain benchmark configurations.

All cases in this section use the nondimensional lattice speed
\begin{equation}
  c=\frac{\Delta x}{\Delta t}=5.
  \label{eq:sec4-lattice-speed}
\end{equation}
The value was selected before the refinement studies and kept fixed under acoustic scaling. Table~\ref{tab:validation-setup} also reports the largest tangent material wave speed observed along each validation trajectory. The acoustic tensor \(Q(\bn;\bF)\) from Section~\ref{subsec:first-order-system} was evaluated over all nodes and time levels of the finest discretization. For the manufactured-solution cases, the audit was performed directly on the prescribed analytical deformation history. For the external benchmarks, it was performed on the deformation-gradient history recorded from the finest \(160^2\) numerical calculation. Since \(Q(\bn;\bF)=Q(-\bn;\bF)\), the directional maximization was approximated by scanning \(\theta\in[0,\pi]\) with an angular increment of \(0.25^\circ\). All measured ratios \(c_{\max}^{\rm tan}/c\) are below one; the largest value is \(0.542\), corresponding to \(c_{\max}^{\rm tan}=2.71<c=5\). Thus the lattice speed used in this section remains above the maximum physical wave speed encountered in each validation case.

The manufactured-solution cases use \(\omega=2\), the nondissipative relaxation limit used in the consistency assessment, because the exact fields are smooth and the target is the formal discretization error. The external tension and shear benchmarks use
\begin{equation}
  \omega=2-1.6\Delta x.
  \label{eq:benchmark-omega}
\end{equation}
The departure \(2-\omega=O(\Delta x)\) adds a small refinement-vanishing lattice dissipation. It damps boundary and loading transients associated with the half-way traction and velocity reconstructions in the ramped benchmark problems, while preserving the continuum hyperelastic benchmark as the target problem.

\begin{table}[pos=htbp]
\centering
\caption{Validation cases in Section~\ref{sec:numerical-validation}. For the benchmark rows, Eq.~\eqref{eq:benchmark-omega} gives \(\omega=1.92,1.96,1.98,1.99\) on \(n=20,40,80,160\). The last column is the trajectory maximum of the tangent material wave-speed ratio.}
\label{tab:validation-setup}
\scriptsize
\setlength{\tabcolsep}{4pt}
\begin{tabular}{@{}p{0.23\textwidth}p{0.14\textwidth}p{0.31\textwidth}p{0.10\textwidth}p{0.09\textwidth}@{}}
\toprule
Case & Reference solution & Material and loading & \(\omega\) & \(c_{\max}^{\rm tan}/c\) \\
\midrule
Periodic manufactured solution & Analytic MMS & log-NH, \(\lambda=2/3\), \(\mu=1\), amplitude \(0.08\) & 2 & 0.542 \\
Boundary manufactured solution & Analytic MMS & log-NH, \(\lambda=2/3\), \(\mu=1\), amplitude \(0.08\) & 2 & 0.542 \\
Uniaxial tension & Q1 FE & SVK, \(\nu=0.20\), \(\alpha=0.175\) & \(2-1.6\Delta x\) & 0.359 \\
Uniaxial tension & Q1 FE & NH, \(\nu=-0.10\), \(\alpha=0.35\) & \(2-1.6\Delta x\) & 0.271 \\
Simple shear & Q1 FE & SVK, \(\nu=0\), \(\alpha=0.03\) & \(2-1.6\Delta x\) & 0.320 \\
Simple shear & Q1 FE & NH, \(\nu=0.20\), \(\alpha=0.10\) & \(2-1.6\Delta x\) & 0.399 \\
\bottomrule
\end{tabular}
\end{table}

\FloatBarrier

Errors are reported in relative \(L^2\) norms over all grid nodes and tensor components. For the manufactured-solution tests we use
\[
  E_{L2}(u)=\frac{\|\bu_h-\bu\|_2}{\|\bu\|_2},\qquad
  E_{L2}(P)=\frac{\|\bP_h-\bP\|_2}{\|\bP\|_2},\qquad
  E_{L2}(\sigma)=\frac{\|\bm\sigma_h-\bm\sigma\|_2}{\|\bm\sigma\|_2}.
\]
For the boundary manufactured solution, the same quantities are also evaluated after removing two grid layers next to the boundary. In the finite-strain benchmark cases of M\"uller et al.~\citep{muller2025nonlinear}, we additionally report the measures defined in that reference,
\[
  E_2=\frac{1}{N_{\rm node}}\frac{\|\bu_h-\bu_{\rm FE}\|_2}{\|\bu_{\rm FE}\|_2},\qquad
  E_\infty=\frac{\|\bu_h-\bu_{\rm FE}\|_\infty}{\|\bu_{\rm FE}\|_2},
\]
where \(N_{\rm node}\) is the total number of lattice nodes and \(\|\cdot\|_\infty\) is the maximum over grid nodes and displacement components. We also retain the raw relative \(L^2\) displacement error \(\|\bu_h-\bu_{\rm FE}\|_2/\|\bu_{\rm FE}\|_2\) for convergence plots; this raw quantity equals \(N_{\rm node}E_2\).

\subsection{Periodic manufactured solution}
\label{subsec:periodic-mms-validation}

The periodic manufactured solution removes boundary effects and directly tests the nonlinear moment matching in Eq.~\eqref{eq:feq}. The reference problem is posed on \(\Omega_0=(0,1)^2\) with periodic boundaries in both material-coordinate directions. The material is the logarithmic compressible neo-Hookean solid listed in Appendix~\ref{app:constitutive-laws}, with \(\lambda=2/3\) and \(\mu=1\). The exact displacement is
\begin{subequations}
\label{eq:periodic-mms-displacement}
\begin{align}
  u_x(X,Y,t)
  &=A\sin\!\bigl(k(X-0.23t)\bigr)
      \cos\!\bigl(k(Y-0.17t)\bigr)
      \sin\!\bigl(k(t-0.11)\bigr),\\
  u_y(X,Y,t)
  &=A\cos\!\bigl(k(X+0.19t)\bigr)
      \sin\!\bigl(k(Y-0.29t)\bigr)
      \cos\!\bigl(k(t+0.07)\bigr),
\end{align}
\end{subequations}
with \(A=0.08\) and \(k=2\pi\). The body acceleration is evaluated from the exact balance,
\begin{equation}
  \bb=\partial_t^2\bu-\Div\bP(\Id+\Grad\bu),
  \label{eq:mms-body-force}
\end{equation}
so that the prescribed \(\bu\) satisfies Eq.~\eqref{eq:displacement-form} exactly. At \(t=0.1\), the deformation is visibly finite: the exact Jacobian ranges from \(0.781\) to \(1.205\), and the maximum value of \(\|\bF-\Id\|\) is \(0.244\). The test therefore exercises the nonlinear stress fluxes in a clearly finite-strain regime.

\begin{figure}[pos=htbp]
\centering
\includegraphics[width=\MainFigureWidth]{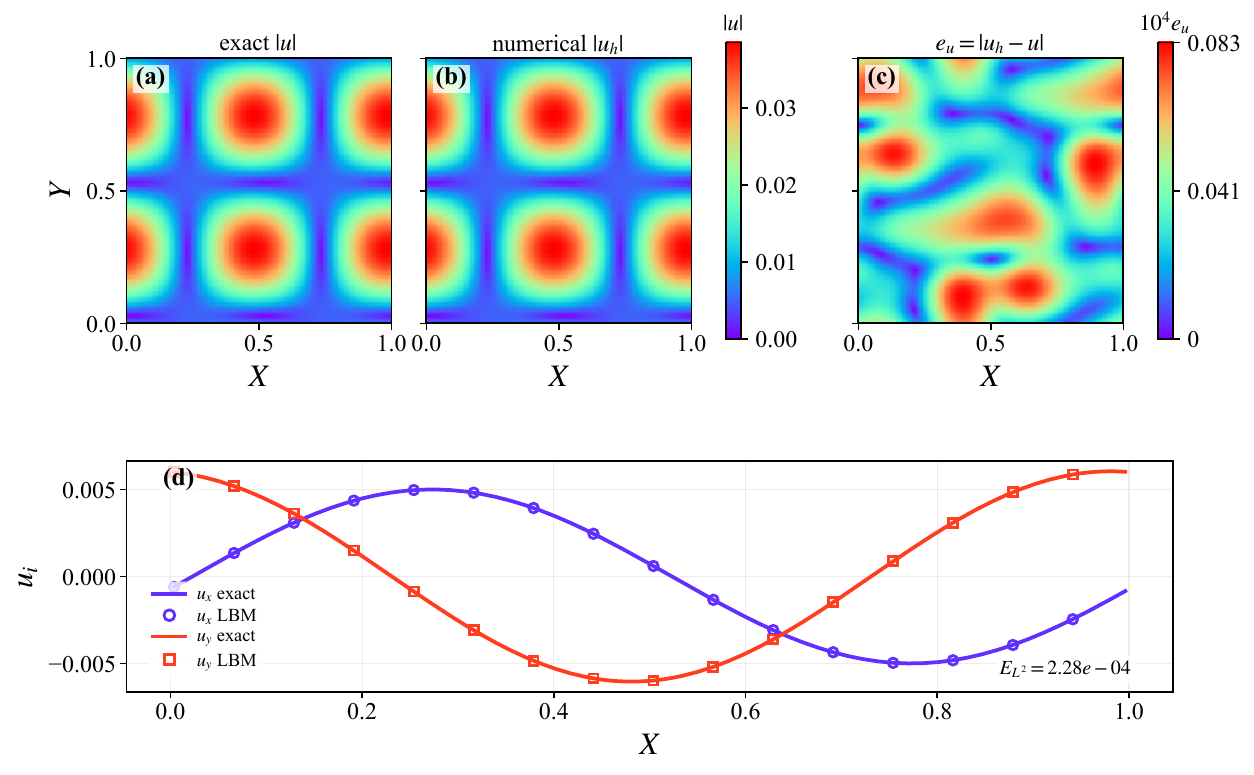}
\caption{Periodic manufactured solution at \(t=0.1\) on a \(128^2\) grid. The panels compare the exact displacement magnitude, the D2Q4\(\times\)6 result, the pointwise error, and a centerline trace.}
\label{fig:periodic-mms-fields}
\end{figure}

Figure~\ref{fig:periodic-mms-fields} compares the exact and numerical displacement magnitude on a \(128^2\) grid, together with the pointwise error and a representative centerline profile. The displacement field, local error pattern, and one-dimensional trace agree closely. The relative errors at this resolution are \(2.28\times10^{-4}\) for displacement, \(5.03\times10^{-4}\) for first Piola stress, and \(5.63\times10^{-4}\) for Cauchy stress; these values are included in the manufactured-solution error summary in Table~\ref{tab:mms-error-summary}.

\begin{figure}[pos=htbp]
\centering
\includegraphics[width=\MainFigureWidth]{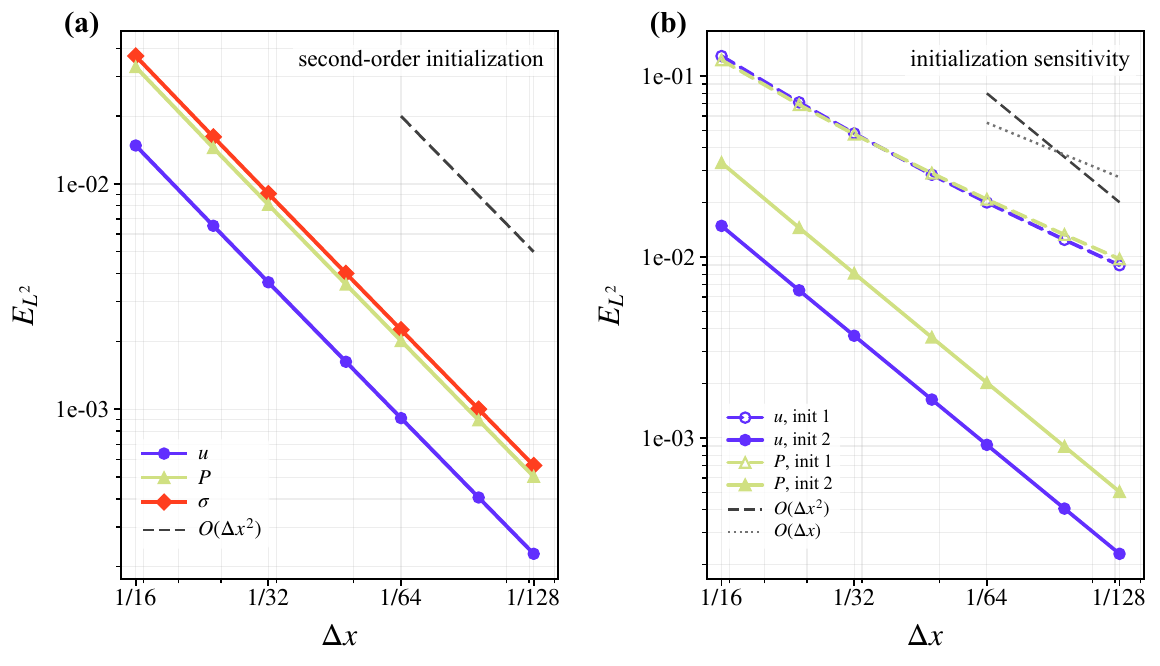}
\caption{Grid convergence for the periodic manufactured solution. The left panel reports second-order initialized errors in \(u\), \(P\), and \(\sigma\). The right panel compares equilibrium initialization with the second-order population initialization.}
\label{fig:periodic-mms-convergence}
\end{figure}

The refinement study in Fig.~\ref{fig:periodic-mms-convergence} separates the spatial-temporal discretization error from the population initialization error. With the second-order initialization of Eq.~\eqref{eq:init-compact}, the grid sequence shows second-order convergence for \(u\), \(P\), and \(\sigma\). An equilibrium-only initialization leaves a short initial kinetic layer and gives approximately first-order behavior over the same time interval, as shown by the open-symbol curves. The stress norms are included because the stress is the nonlinear flux carried by the lattice moments and is more sensitive to errors in \(\bF\) than displacement alone.

\FloatBarrier

\subsection{Manufactured solution with boundary data}
\label{subsec:boundary-mms-validation}

The second manufactured-solution test uses the same nonlinear reference field in Eq.~\eqref{eq:periodic-mms-displacement}, logarithmic neo-Hookean material, and body acceleration in Eq.~\eqref{eq:mms-body-force}, now on the bounded square \(\Omega_0=(0,1)^2\). Two boundary configurations are considered. The all-Dirichlet case prescribes the exact material velocity on each side,
\begin{equation}
  \bv=\partial_t\bu
  \qquad \text{on }\partial\Omega_0.
  \label{eq:boundary-mms-dirichlet}
\end{equation}
The mixed case imposes exact velocity conditions on the left and bottom sides and exact first-Piola traction on the right and top sides,
\begin{equation}
  \bv=\partial_t\bu
  \quad \text{on }X=0\text{ and }Y=0,
  \qquad
  \bP\bN=\bP(\Id+\Grad\bu)\bN
  \quad \text{on }X=1\text{ and }Y=1.
  \label{eq:boundary-mms-mixed}
\end{equation}
The mixed case exercises the traction reconstruction in Eq.~\eqref{eq:neumann-boundary-rule}, including the local Newton solve for the boundary deformation gradient.

\begin{figure}[pos=htbp]
\centering
\includegraphics[width=\MainFigureWidth]{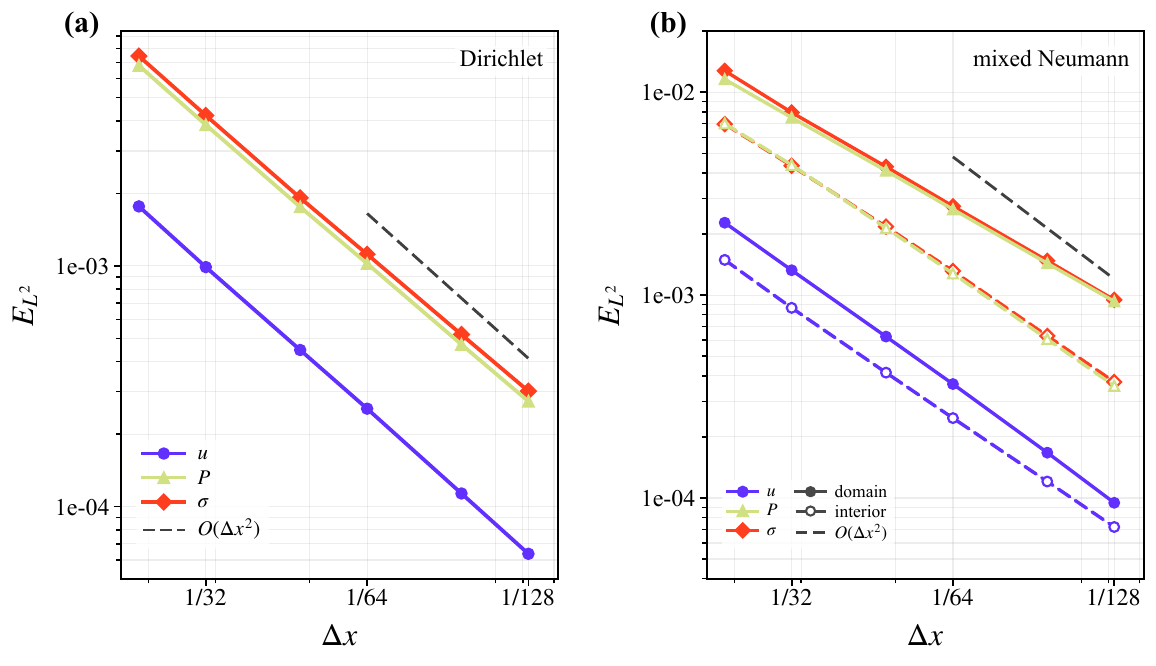}
\caption{Manufactured solution with boundary data. The figure compares all-Dirichlet boundary conditions with mixed Dirichlet--Neumann boundary conditions and reports both full-domain and two-layer-stripped interior errors.}
\label{fig:boundary-mms-convergence}
\end{figure}

Figure~\ref{fig:boundary-mms-convergence} shows the full-domain and interior convergence histories. The all-Dirichlet case gives second-order convergence in displacement and stress. At \(n=128\), the full-domain relative errors are \(6.37\times10^{-5}\), \(2.74\times10^{-4}\), and \(3.02\times10^{-4}\) for \(u\), \(P\), and \(\sigma\). Removing two boundary layers gives \(5.82\times10^{-5}\), \(2.28\times10^{-4}\), and \(2.51\times10^{-4}\).

The mixed Neumann case has larger stress errors near the traction boundary, as expected for a nonlinear boundary inversion applied on a half-way lattice link. The interior errors remain systematically smaller than the full-domain values. At \(n=128\), the full-domain relative errors are \(9.47\times10^{-5}\), \(9.31\times10^{-4}\), and \(9.47\times10^{-4}\), while the corresponding interior errors are \(7.19\times10^{-5}\), \(3.57\times10^{-4}\), and \(3.73\times10^{-4}\). The interior convergence confirms second-order bulk accuracy of the nonlinear flux reconstruction, with the full-domain stress norms retaining a localized boundary contribution. Table~\ref{tab:mms-error-summary} summarizes the finest-grid errors for both manufactured-solution tests.

\FloatBarrier

\begin{table}[pos=htbp]
\centering
\caption{Representative relative errors from the manufactured-solution validation cases at the finest grid used in each study. Boundary interior errors remove two grid layers adjacent to the boundary.}
\label{tab:mms-error-summary}
\small
\begin{tabular}{@{}llccc@{}}
\toprule
Case & Region or initialization & \(E_{L2}(u)\) & \(E_{L2}(P)\) & \(E_{L2}(\sigma)\) \\
\midrule
Periodic MMS, \(n=128\) & second-order initialization & \(2.28\times10^{-4}\) & \(5.03\times10^{-4}\) & \(5.63\times10^{-4}\) \\
Periodic MMS, \(n=128\) & equilibrium initialization & \(8.97\times10^{-3}\) & \(9.72\times10^{-3}\) & \(9.75\times10^{-3}\) \\
Dirichlet boundary MMS, \(n=128\) & full domain & \(6.37\times10^{-5}\) & \(2.74\times10^{-4}\) & \(3.02\times10^{-4}\) \\
Dirichlet boundary MMS, \(n=128\) & interior & \(5.82\times10^{-5}\) & \(2.28\times10^{-4}\) & \(2.51\times10^{-4}\) \\
Mixed Neumann MMS, \(n=128\) & full domain & \(9.47\times10^{-5}\) & \(9.31\times10^{-4}\) & \(9.47\times10^{-4}\) \\
Mixed Neumann MMS, \(n=128\) & interior & \(7.19\times10^{-5}\) & \(3.57\times10^{-4}\) & \(3.73\times10^{-4}\) \\
\bottomrule
\end{tabular}
\end{table}

\FloatBarrier

\subsection{Uniaxial tension benchmark}
\label{subsec:tension-benchmark}

The first external benchmark is the uniaxial-tension problem used by M\"uller et al.~\citep{muller2025nonlinear}. It is posed on \(\Omega_0=(0,1)^2\) with an initially undeformed body at rest. The shear modulus is fixed to \(\mu=1\), and the Lam\'e parameter is set from the two-dimensional Poisson ratio by \(\lambda=2\mu\nu/(1-2\nu)\). The two material cases use the SVK and compressible neo-Hookean laws listed in Appendix~\ref{app:constitutive-laws}. The top and bottom boundaries carry opposite nominal tractions,
\begin{equation}
  \bP\bN=(0,g_\alpha(t))^T \quad \text{on }Y=1,
  \qquad
  \bP\bN=(0,-g_\alpha(t))^T \quad \text{on }Y=0,
  \qquad
  \bP\bN=\bzero \quad \text{on }X=0,1,
  \label{eq:tension-benchmark-bc}
\end{equation}
where
\begin{equation}
  g_\alpha(t)=
  \begin{cases}
    \alpha\sin^2(\pi t/4), & 0\le t<2,\\
    \alpha, & t\ge 2.
  \end{cases}
  \label{eq:tension-load}
\end{equation}
We consider the SVK case with \(\nu=0.20\), \(\alpha=0.175\), and the compressible neo-Hookean case with \(\nu=-0.10\), \(\alpha=0.35\). The final time is \(t=2.2\). The numerical displacement is compared against an independently generated Q1 finite-element reference sampled at the lattice nodes. The same displacement field is then used to compute the \(E_2\) and \(E_\infty\) benchmark metrics.

\begin{figure}[pos=htbp]
\centering
\includegraphics[width=\MainFigureWidth]{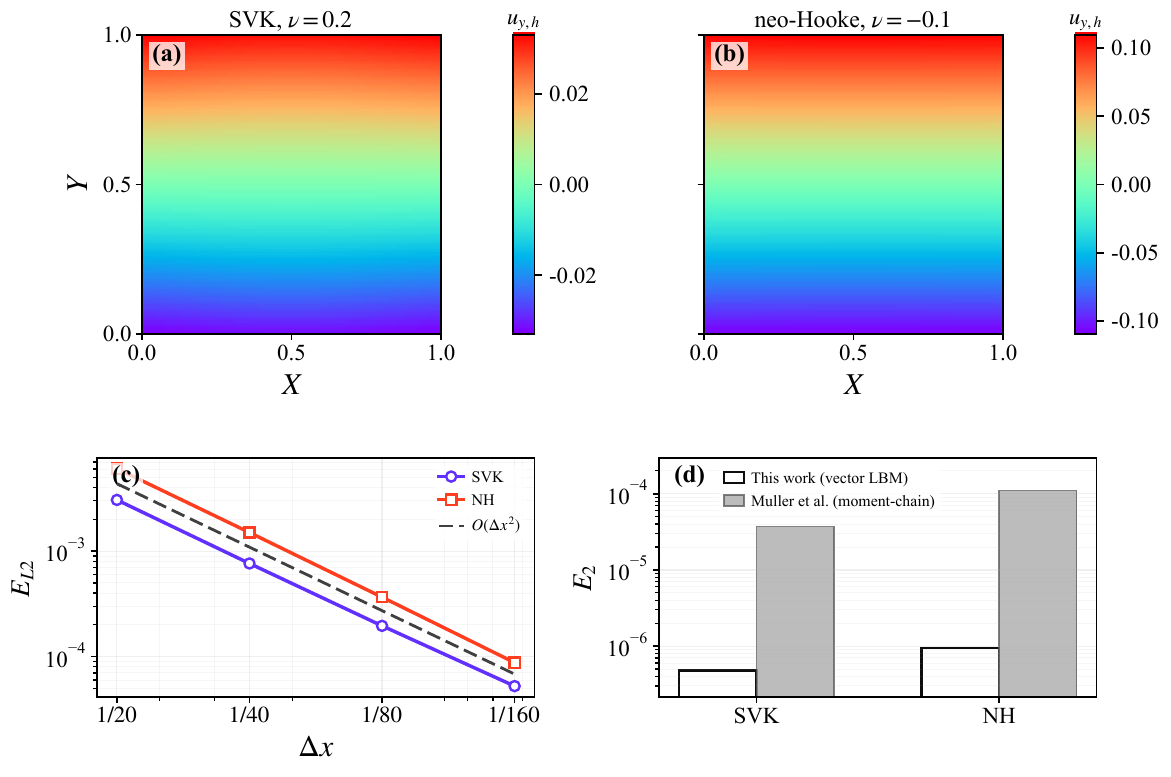}
\caption{Uniaxial-tension benchmark of M\"uller et al.~\citep{muller2025nonlinear}. The panels show the final deformation, displacement error against the Q1 finite-element reference, raw relative \(L^2\) convergence, and \(E_2\) comparison at the reference-study grid resolution \(n=40\) (\(N_{\rm node}=1600\)).}
\label{fig:tension-benchmark}
\end{figure}

Figure~\ref{fig:tension-benchmark} shows the deformed shape, the reference-error field, and the grid convergence for the two material choices. The raw relative \(L^2\) displacement error shows second-order convergence for the vectorial lattice scheme. On the \(160^2\) grid, the raw relative errors are \(5.25\times10^{-5}\) for the SVK case and \(8.77\times10^{-5}\) for the neo-Hookean case. The comparison with the published moment-chain data of M\"uller et al.~\citep{muller2025nonlinear} is restricted to the matching reference-study grid resolution, \(n=40\) (\(N_{\rm node}=1600\)). At this resolution, the present \(E_2\) values are \(4.78\times10^{-7}\) and \(9.39\times10^{-7}\), lower than the reported \(3.7\times10^{-5}\) and \(1.1\times10^{-4}\) values by factors of \(7.8\times10^1\) and \(1.2\times10^2\), respectively. The corresponding \(E_\infty\) values are summarized in Table~\ref{tab:external-benchmark-summary}.

\subsection{Simple shear benchmark}
\label{subsec:simple-shear-benchmark}

The second external benchmark is the simple-shear problem from the same reference. It uses the same SVK and compressible neo-Hookean laws and the same initially undeformed state on \(\Omega_0=(0,1)^2\). The lateral boundaries are traction free, the bottom boundary is fixed in velocity, and the top boundary is driven tangentially:
\begin{equation}
  \bP\bN=\bzero \quad \text{on }X=0,1,
  \qquad
  \bv=\bzero \quad \text{on }Y=0,
  \qquad
  \bv=(h_\alpha(t),0)^T \quad \text{on }Y=1,
  \label{eq:shear-benchmark-bc}
\end{equation}
with
\begin{equation}
  h_\alpha(t)=
  \begin{cases}
    \alpha\sin(\pi t/2), & 0\le t<2,\\
    0, & t\ge 2.
  \end{cases}
  \label{eq:shear-velocity}
\end{equation}
The corresponding top displacement is \(u_x=2\alpha[1-\cos(\pi t/2)]/\pi\) during the ramp and \(u_x=4\alpha/\pi\) after \(t=2\). This case probes the coupling between shear deformation, normal stress response, and traction-driven boundary evolution. We use the SVK case with \(\nu=0\) and \(\alpha=0.03\), and the neo-Hookean case with \(\nu=0.20\) and \(\alpha=0.10\). As in the tension benchmark, all errors are measured against an independent Q1 finite-element reference at \(t=2.2\).

\begin{figure}[pos=htbp]
\centering
\includegraphics[width=\MainFigureWidth]{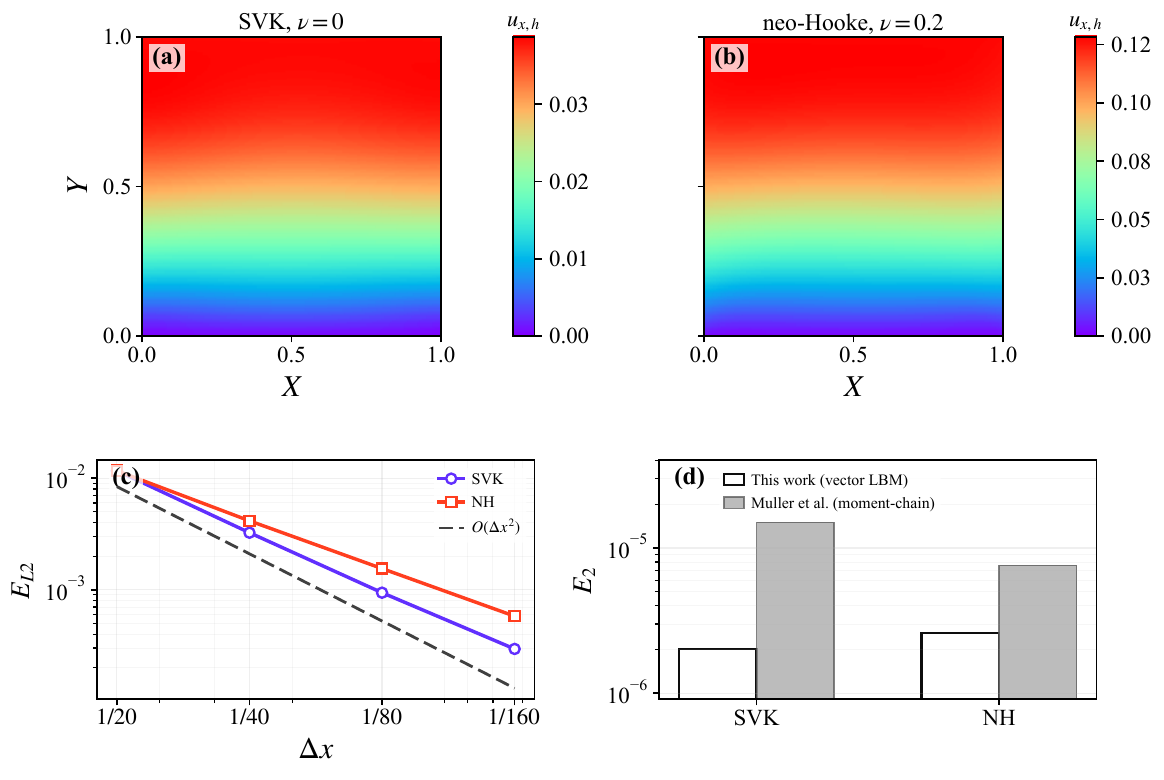}
\caption{Simple-shear benchmark of M\"uller et al.~\citep{muller2025nonlinear}. The figure reports final deformation, displacement error against the Q1 finite-element reference, raw relative \(L^2\) convergence, and \(E_2\) comparison at the reference-study grid resolution \(n=40\) (\(N_{\rm node}=1600\)).}
\label{fig:simple-shear-benchmark}
\end{figure}

Figure~\ref{fig:simple-shear-benchmark} shows that the vectorial formulation follows the finite-element displacement field over the full loading interval. The convergence has larger error constants than in uniaxial tension, reflecting the stronger shear-driven boundary gradients in this setup. At \(n=160\), the raw relative displacement errors are \(2.95\times10^{-4}\) for the SVK case and \(5.80\times10^{-4}\) for the neo-Hookean case. As above, the moment-chain comparison is restricted to the matching reference-study grid resolution, \(n=40\). At this resolution, the present \(E_2\) values are \(2.03\times10^{-6}\) and \(2.59\times10^{-6}\), lower than the reported \(1.5\times10^{-5}\) and \(7.6\times10^{-6}\) values by factors of \(7.4\) and \(2.9\), respectively. The benchmark confirms that the same boundary and flux reconstruction used in the manufactured tests carries over to a reference finite-strain shear configuration.

\FloatBarrier

The benchmark comparisons at the reference-study grid resolution are summarized in Table~\ref{tab:external-benchmark-summary}.

\begin{table}[pos=htbp]
\centering
\caption{Benchmark comparison with the published moment-chain data of M\"uller et al.~\citep{muller2025nonlinear} at \(n=40\) (\(N_{\rm node}=1600\)). The quantities \(E_2\) and \(E_\infty\) follow the benchmark definitions in that reference.}
\label{tab:external-benchmark-summary}
\footnotesize
\setlength{\tabcolsep}{5pt}
\begin{tabular}{@{}llccccc@{}}
\toprule
Benchmark & Material case & \(n\) & \(E_2\) & \(E_2\) ref. & \(E_\infty\) & \(E_\infty\) ref. \\
\midrule
Uniaxial tension & SVK, \(\nu=0.20\) & 40 & \(4.78\times10^{-7}\) & \(3.7\times10^{-5}\) & \(9.40\times10^{-5}\) & \(5.7\times10^{-3}\) \\
Uniaxial tension & NH, \(\nu=-0.10\) & 40 & \(9.39\times10^{-7}\) & \(1.1\times10^{-4}\) & \(1.05\times10^{-4}\) & \(1.8\times10^{-2}\) \\
Simple shear & SVK, \(\nu=0\) & 40 & \(2.03\times10^{-6}\) & \(1.5\times10^{-5}\) & \(2.70\times10^{-4}\) & \(1.6\times10^{-3}\) \\
Simple shear & NH, \(\nu=0.20\) & 40 & \(2.59\times10^{-6}\) & \(7.6\times10^{-6}\) & \(5.94\times10^{-4}\) & \(1.0\times10^{-3}\) \\
\bottomrule
\end{tabular}
\end{table}

\FloatBarrier

Overall, the manufactured-solution and external-benchmark results support the numerical accuracy of the D2Q4\(\times\)6 construction on periodic, Dirichlet, and mixed traction boundaries. They also show that the same flux and boundary reconstructions carry from controlled reference fields to finite-strain tension and shear benchmark configurations.

\section{Constitutive response and finite-strain wave dynamics}
\label{sec:constitutive-wave-dynamics}

The preceding validation cases establish the accuracy of the nonlinear flux discretization and the grid-aligned boundary closures. We now examine the same formulation in settings chosen to isolate material response, tangent dynamics about finite deformations, finite-amplitude wave propagation, and a bounded-domain bending wave. The first three cases use analytical or semi-analytical references. The final case uses an independent finite-element reference and provides a longer-time dynamic assessment on a cantilever geometry.

\subsection{Affine finite-strain material response}
\label{subsec:affine-patch-suite}

A homogeneous deformation gives a direct view of the constitutive interface used by the lattice method. Consider an affine motion
\begin{equation}
  \bx(\bX,t)=\bF(t)\bX+\bm\beta(t),
  \qquad
  \bv(\bX,t)=\dot{\bF}(t)\bX+\dot{\bm\beta}(t),
  \label{eq:affine-patch-motion}
\end{equation}
where \(\bm\beta(t)\) is a time-dependent rigid translation and overdots denote time derivatives. The deformation gradient is spatially constant. Since \(\bP(\bF(t))\) is also spatially constant, \(\Div\bP=\bzero\). In the numerical calculations \(\dot{\bF}\) is constant during the loading interval, so the affine velocity boundary data generate an exact bulk solution without body forcing. This construction leaves the constitutive law free: any admissible \(W(\bF)\) in Appendix~\ref{app:constitutive-laws} supplies an exact stress path.

\begin{figure}[pos=htbp]
\centering
\includegraphics[width=\textwidth]{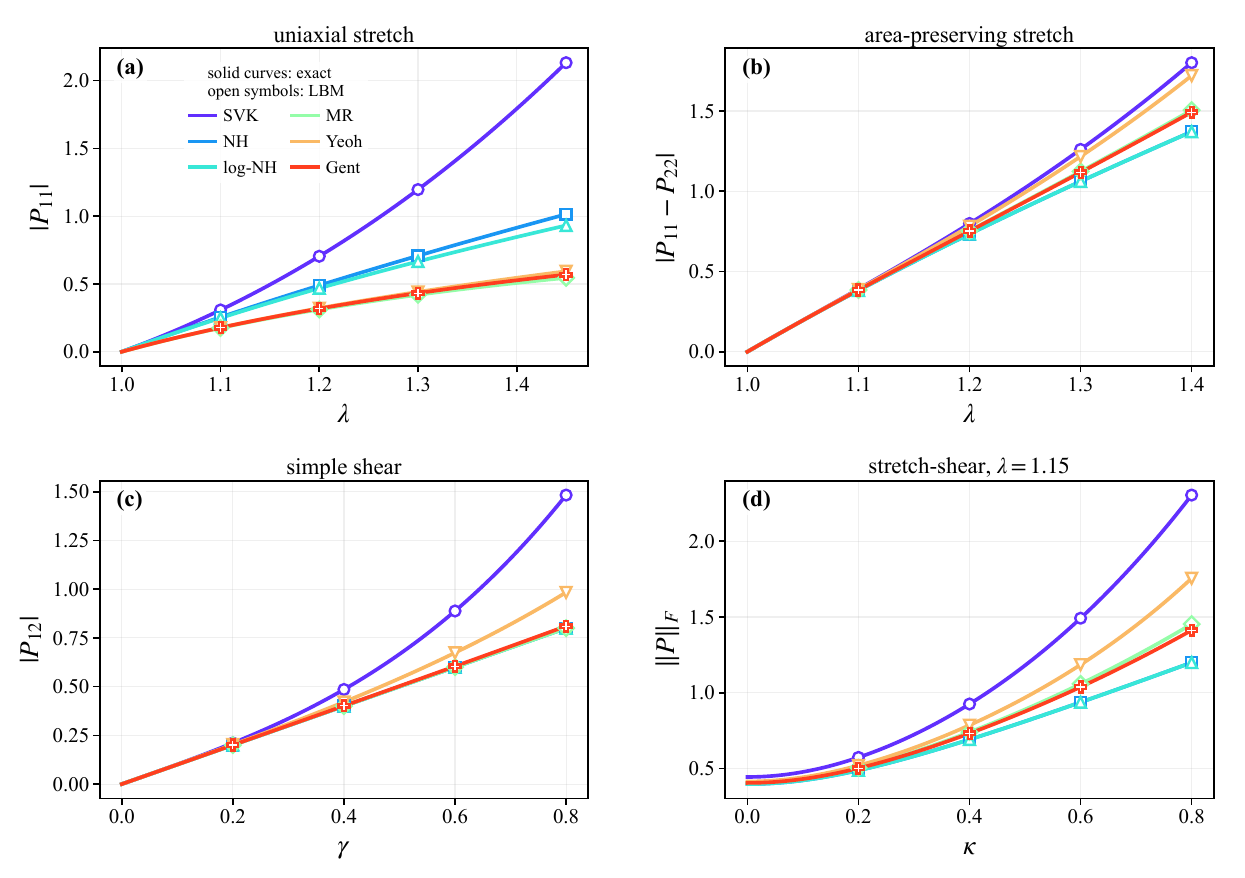}
\caption{Constitutive stress-response curves for homogeneous affine finite-strain paths. Solid curves are analytical stresses from the constitutive laws in Appendix~\ref{app:constitutive-laws}; open symbols are D2Q4\(\times\)6 affine-patch samples on an \(80^2\) grid.}
\label{fig:constitutive-stress-curves}
\end{figure}

The affine-patch calculations set \(\bm\beta(t)=\bzero\), because rigid translation does not affect \(\bF\), \(\bP\), or \(\bm\sigma\). Over a loading interval \(0\le t\le T\), the exact reference fields are
\begin{equation}
  \bF(t)=\Id+\frac{t}{T}(\bF_*-\Id),\qquad
  \bu(\bX,t)=\bigl[\bF(t)-\Id\bigr]\bX,\qquad
  \bv(\bX,t)=\frac{\bF_*-\Id}{T}\,\bX,
  \label{eq:affine-exact-fields}
\end{equation}
with stresses obtained pointwise as \(\bP(t)=\bP(\bF(t))\) and \(\bm\sigma(t)=J^{-1}\bP(t)\bF(t)^T\). The stress curves in Fig.~\ref{fig:constitutive-stress-curves} use the terminal states \(\bF_*\) from the following four families:
\begin{equation}
  \bF_{*,{\rm uni}}=
  \begin{pmatrix}
    \lambda_s & 0\\
    0 & 1
  \end{pmatrix},
  \quad
  \bF_{*,{\rm aps}}=
  \begin{pmatrix}
    \lambda_s & 0\\
    0 & \lambda_s^{-1}
  \end{pmatrix},
  \quad
  \bF_{*,{\rm sh}}=
  \begin{pmatrix}
    1 & \gamma\\
    0 & 1
  \end{pmatrix},
  \quad
  \bF_{*,{\rm ss}}=
  \begin{pmatrix}
    1.15 & \kappa\\
    0 & 1/1.15
  \end{pmatrix}.
  \label{eq:affine-deformation-paths}
\end{equation}
They represent uniaxial stretch, area-preserving stretch, simple shear, and combined stretch--shear. Here \(\lambda_s\) is the scalar stretch parameter, \(\gamma\) is the simple-shear parameter, and \(\kappa\) is the shear parameter added to the fixed area-preserving stretch with axial stretch \(1.15\). In the stress-response curves, \(\lambda_s\in[1,1.45]\) for uniaxial stretch, \(\lambda_s\in[1,1.40]\) for area-preserving stretch, and \(\gamma,\kappa\in[0,0.80]\). The material set consists of SVK, compressible neo-Hookean, logarithmic neo-Hookean, Mooney--Rivlin, Yeoh, and Gent solids, all evaluated through the same pointwise map \(\bP(\bF)\).

Figure~\ref{fig:constitutive-stress-curves} compares analytical stress-response curves with numerical affine-patch samples. The selected scalar response is \(P_{11}\) for uniaxial stretch, \(P_{11}-P_{22}\) for area-preserving stretch, \(P_{12}\) for simple shear, and \(\|\bP\|_F\), the Frobenius norm of \(\bP\), for the combined stretch--shear path. The numerical markers lie on the corresponding analytical curves across the six material laws. Thus the same lattice update follows distinct finite-strain stress responses through the local evaluation of \(\bP(\bF)\).

\begin{figure}[pos=htbp]
\centering
\includegraphics[width=\textwidth]{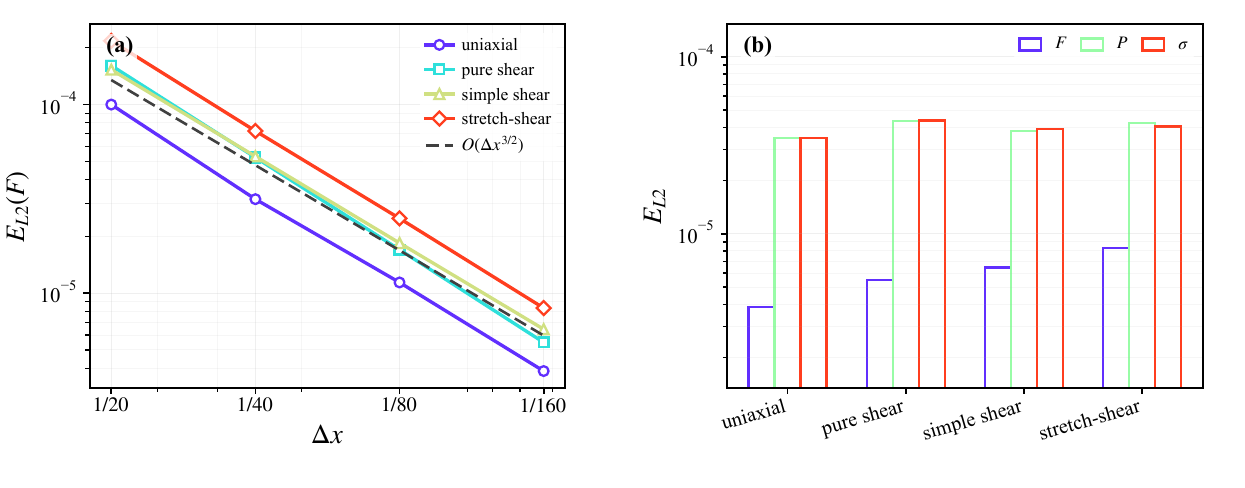}
\caption{Affine finite-strain patch errors. Panel (a) reports grid refinement of the deformation-gradient error for the four affine paths. Panel (b) summarizes the \(n=160\) relative errors in \(\bF\), \(\bP\), and \(\bm\sigma\), taking the maximum over the six material laws for each path.}
\label{fig:affine-patch-errors}
\end{figure}

The corresponding tensor errors are shown in Fig.~\ref{fig:affine-patch-errors}. The refinement study uses \(n=20,40,80,160\), \(c=5\), second-order initialization, and the same refinement-scaled relaxation used in the external benchmarks. The deformation-gradient error decreases systematically for all four paths, with observed slopes between approximately \(1.5\) and \(1.7\) over the grid sequence. On the \(160^2\) grid, the largest relative errors over all material--path combinations are \(8.34\times10^{-6}\) in \(\bF\), \(4.34\times10^{-5}\) in \(\bP\), and \(4.36\times10^{-5}\) in \(\bm\sigma\). The stress errors therefore follow the directly evolved deformation gradient, confirming that the post-processed Piola and Cauchy stresses remain consistent with the affine exact solution.

\subsection{Wave speeds about predeformed states}
\label{subsec:predeformed-acoustic-wave-suite}

The next case probes the tangent dynamics about finite homogeneous states. Let \(\bF_0\) be a constant predeformation and let a small displacement perturbation travel in the material direction \(\bn\). Using the acoustic tensor introduced in Eq.~\eqref{eq:acoustic-tensor-section2}, the analytical phase speeds are obtained from
\begin{equation}
  Q(\bn;\bF_0)\bm q=c_{\rm ac}^2\bm q,
  \qquad\text{equivalently}\qquad
  Q_{ik}(\bn;\bF_0)q_k=c_{\rm ac}^2q_i,
  \label{eq:predeformed-acoustic-eigenproblem}
\end{equation}
where \(\bm q=(q_1,q_2)^T\) is the polarization vector, \(q_i\) denotes its \(i\)-th component, and \(c_{\rm ac}\) is the tangent wave speed. The prescribed finite predeformations are then sampled directly through periodic wave propagation.

For a small perturbation amplitude \(\epsilon\) and wavenumber \(k\), the corresponding linearized reference field is
\begin{equation}
  \bu'(\bX,t)=\epsilon\,\bm q\sin\!\bigl(k\,\bn\cdot\bX-kc_{\rm ac}t\bigr),\qquad
  \bv'(\bX,t)=-\epsilon kc_{\rm ac}\bm q\cos\!\bigl(k\,\bn\cdot\bX-kc_{\rm ac}t\bigr),
  \label{eq:predeformed-acoustic-reference}
\end{equation}
and
\begin{equation}
  \bF(\bX,t)=\bF_0+\epsilon k\,\bm q\otimes\bn\,
  \cos\!\bigl(k\,\bn\cdot\bX-kc_{\rm ac}t\bigr)+O(\epsilon^2).
  \label{eq:predeformed-acoustic-F}
\end{equation}
The phase-speed comparison below uses the material direction \(\bn=(1,0)^T\), \(k=2\pi\), and \(\epsilon=10^{-4}\).

The computations use periodic domains, \(n=80\), \(c=5\), and \(\omega=2\). Two predeformations are considered,
\[
  \bF_0=
  \begin{pmatrix}
    1.20 & 0\\
    0 & 1/1.20
  \end{pmatrix}
  \quad\text{and}\quad
  \bF_0=
  \begin{pmatrix}
    1 & 0.30\\
    0 & 1
  \end{pmatrix},
\]
corresponding to area-preserving stretch and simple shear. For each material and each \(\bF_0\), the slow and fast acoustic branches are initialized with their eigenvectors. The numerical phase speed is then measured from the phase shift of the projected velocity field.
Specifically, the simulated velocity is projected onto the polarization,
\[
  v_q(X,t)=\bm q\cdot\bv(X,t),
\]
and the fundamental Fourier mode is fitted as
\[
  a_c(t)=2\langle v_q(\cdot,t)\cos(kX)\rangle,\qquad
  a_s(t)=2\langle v_q(\cdot,t)\sin(kX)\rangle,\qquad
  \phi(t)=\operatorname{atan2}\bigl(a_s(t),a_c(t)\bigr).
\]
Here \(\operatorname{atan2}\) denotes the two-argument arctangent.
After unwrapping the phase relative to the initial value, the measured lattice phase speed is \(c_{\rm num}=\Delta\phi/(kt)\).

\begin{figure}[pos=htbp]
\centering
\includegraphics[width=\textwidth]{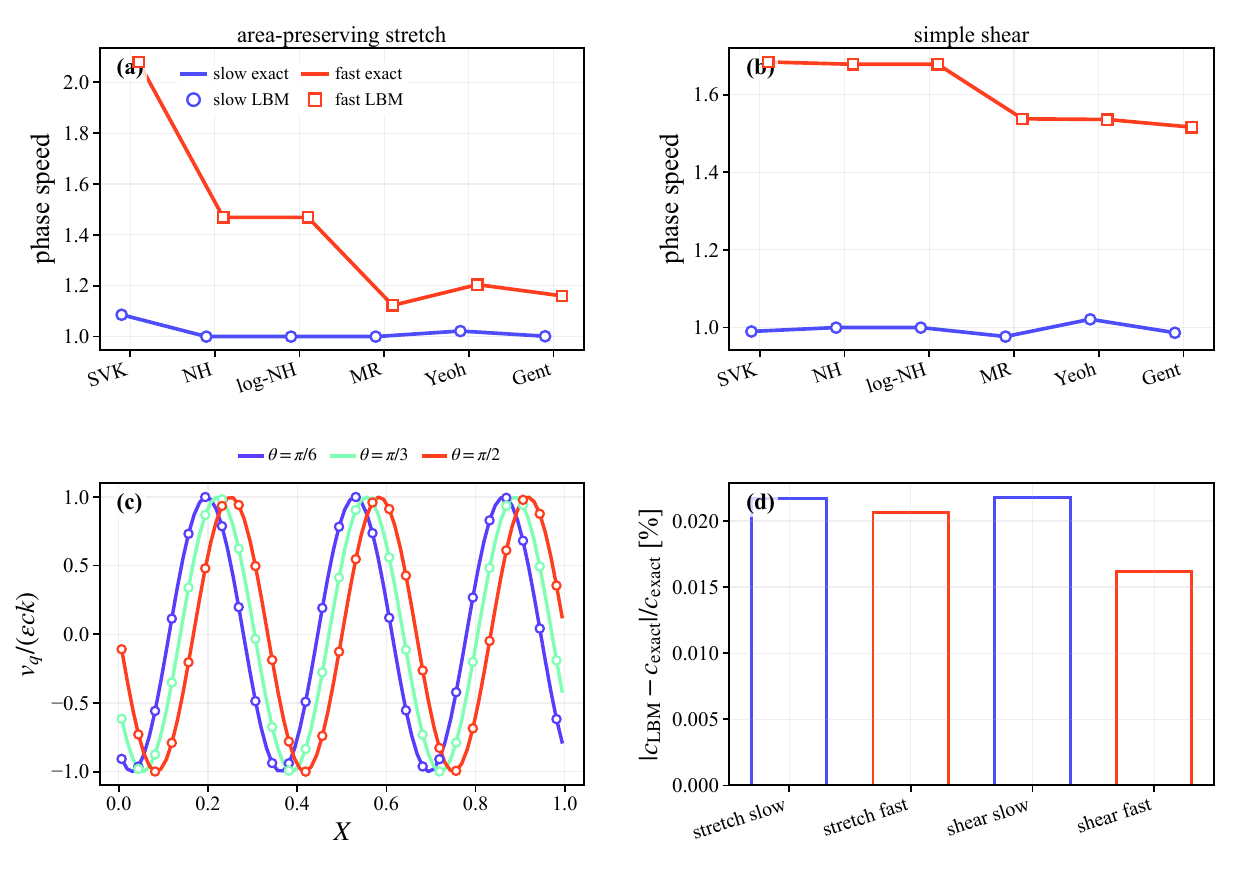}
\caption{Wave speeds about predeformed states. Analytical speeds are obtained from the acoustic tensor \(Q(\bn;\bF_0)\), and open symbols denote phase speeds measured from D2Q4\(\times\)6 simulations. The two branches correspond to the slow and fast eigenmodes of Eq.~\eqref{eq:predeformed-acoustic-eigenproblem}.}
\label{fig:predeformed-acoustic-wave}
\end{figure}

Figure~\ref{fig:predeformed-acoustic-wave} shows the comparison between acoustic-tensor speeds and lattice phase speeds. Across the six material laws, two predeformations, and both branches, the maximum relative speed difference is \(2.18\times10^{-4}\). The agreement demonstrates that the finite predeformation enters the measured dynamics through the current tangent moduli.

\subsection{Finite-amplitude periodic shear wave}
\label{subsec:finite-amplitude-shear-wave}

The acoustic-wave calculation uses an infinitesimal perturbation. A complementary periodic reference is available for a finite-amplitude shear wave in the compressible neo-Hookean and logarithmic neo-Hookean laws. Let
\begin{equation}
  u_1=0,\qquad
  u_2(X,t)=A\sin(kX-c_sk t),
  \qquad
  c_s=\sqrt{\mu},
  \label{eq:finite-amplitude-shear-displacement}
\end{equation}
where \(A\) is the displacement amplitude, \(k\) is the wavenumber, and \(c_s\) is the shear-wave speed. The calculations below use the unit periodic domain with \(k=4\pi\); the reported shear amplitude is the product \(Ak\). This gives
\begin{equation}
  \bF=
  \begin{pmatrix}
    1 & 0\\
    Ak\cos(kX-c_sk t) & 1
  \end{pmatrix},
  \qquad
  J=1.
  \label{eq:finite-amplitude-shear-F}
\end{equation}
The nonzero velocity component is
\begin{equation}
  v_2(X,t)=-A c_s k\cos(kX-c_sk t).
  \label{eq:finite-amplitude-shear-velocity}
\end{equation}
Along this path the volumetric contribution vanishes and both constitutive laws give
\begin{equation}
  P_{21}(X,t)=\mu F_{21}(X,t)=\mu Ak\cos(kX-c_sk t).
  \label{eq:finite-amplitude-shear-stress}
\end{equation}
The field in Eq.~\eqref{eq:finite-amplitude-shear-displacement} is therefore an exact travelling-wave solution on a periodic domain.

\begin{figure}[pos=!htbp]
\centering
\includegraphics[width=\textwidth]{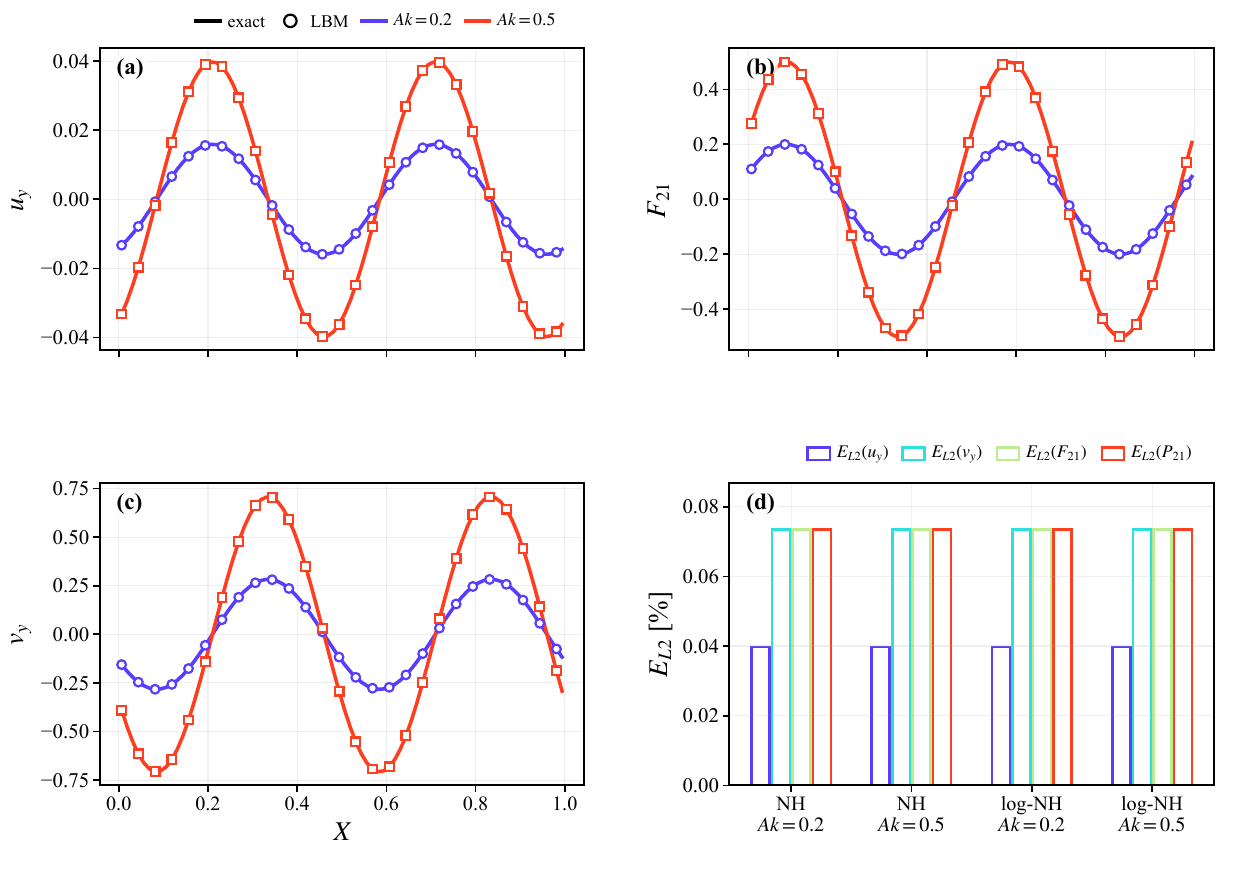}
\caption{Finite-amplitude periodic shear wave for compressible neo-Hookean and logarithmic neo-Hookean materials. Solid curves are the exact travelling-wave solution; open symbols are D2Q4\(\times\)6 results for shear amplitudes \(Ak=0.2\) and \(Ak=0.5\).}
\label{fig:finite-amplitude-shear-wave}
\end{figure}

Figure~\ref{fig:finite-amplitude-shear-wave} compares the exact and numerical profiles for \(\mu=2\), \(n=80\), \(c=5\), \(\omega=2\), and shear amplitudes \(Ak=0.2\) and \(Ak=0.5\). The figure reports the displacement, velocity, deformation-gradient, and Piola-stress components that participate in the wave. The largest relative \(L^2\) error over the plotted cases is \(7.36\times10^{-4}\). The same level of agreement is obtained for the neo-Hookean and logarithmic neo-Hookean closures, as expected from their identical response on this isochoric shear path with \(J=1\) and finite shear component \(F_{21}\).

\FloatBarrier

\subsection{Cantilever bending-wave benchmark}
\label{subsec:cantilever-bending-wave}

The final example considers a bounded-domain wave problem on a cantilever beam. The reference domain is \([0,4]\times[0,1]\), discretized by \(160\times40\) lattice nodes. The left edge is clamped, the top and bottom edges are traction free, and the right edge receives the vertical nominal traction
\[
  \bar{\bT}(Y,t)=\bigl(0,p_R(t)w(Y)\bigr)^T,
  \qquad
  p_R(t)=\frac{2p_0}{\sqrt{3s}\,\pi^{1/4}}(1-\zeta^2)e^{-\zeta^2/2},
  \qquad
  \zeta=\frac{t-t_0}{s},
\]
for \(0\le t\le2\), and \(\bar{\bT}=\bzero\) afterwards. Here \(p_0=0.15\), \(s=0.215\), and \(t_0=1\) are the Ricker-pulse amplitude, width, and centre time. The factor \(w(Y)\) is a smooth right-edge taper that vanishes over eight lattice nodes near the loaded-end corners and is normalized to preserve the resultant load. The material is compressible neo-Hookean with \(\mu=1\) and \(\nu=0.20\). The calculation uses \(c=10\), \(\tau=0.5025\), equivalently \(\omega=1.99005\), where \(\tau=1/\omega\) is the BGK relaxation time, and second-order initialization.

\begin{figure}[pos=!htbp]
\centering
\includegraphics[width=\StackedFigureWidth]{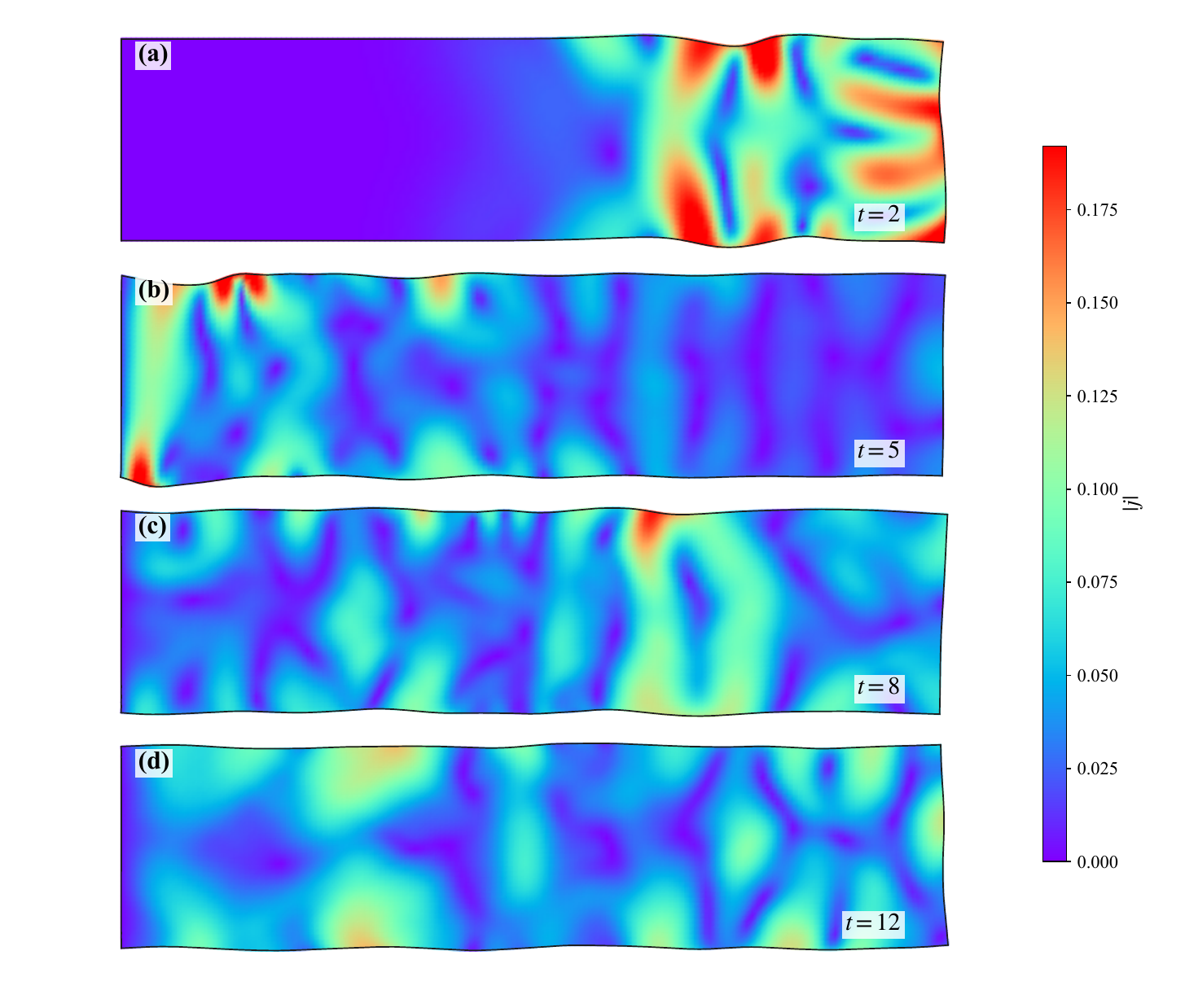}
\caption{Transient bending waves in the \(160\times40\) neo-Hookean cantilever. The panels show the deformed configuration at \(t=2,5,8,12\), colored by the velocity magnitude.}
\label{fig:cantilever-wave-snapshots}
\end{figure}

Figure~\ref{fig:cantilever-wave-snapshots} shows the deformed beam at \(t=2,5,8,12\), colored by the velocity magnitude. The initial pulse enters from the loaded end, reflects within the beam, and produces a bending-wave response over the full integration interval. At \(t=12\), the simulation remains finite with \(0.949\le J\le1.043\). The maximum absolute tip displacement in the vertical direction is \(1.54\times10^{-2}\), while the maximum displacement magnitude over the beam is \(2.81\times10^{-2}\). These values indicate a small but clearly resolved finite-strain wave response on a nonperiodic domain with mixed boundary conditions.

For quantitative assessment, the same cantilever calculation is compared with an independent Q1 finite-element reference. The comparison uses the \(E_2\) and \(E_\infty\) measures introduced in Section~\ref{sec:numerical-validation} for the external benchmarks, evaluated over the full lattice domain.

\begin{figure}[pos=tbp]
\centering
\includegraphics[width=\StackedFigureWidth]{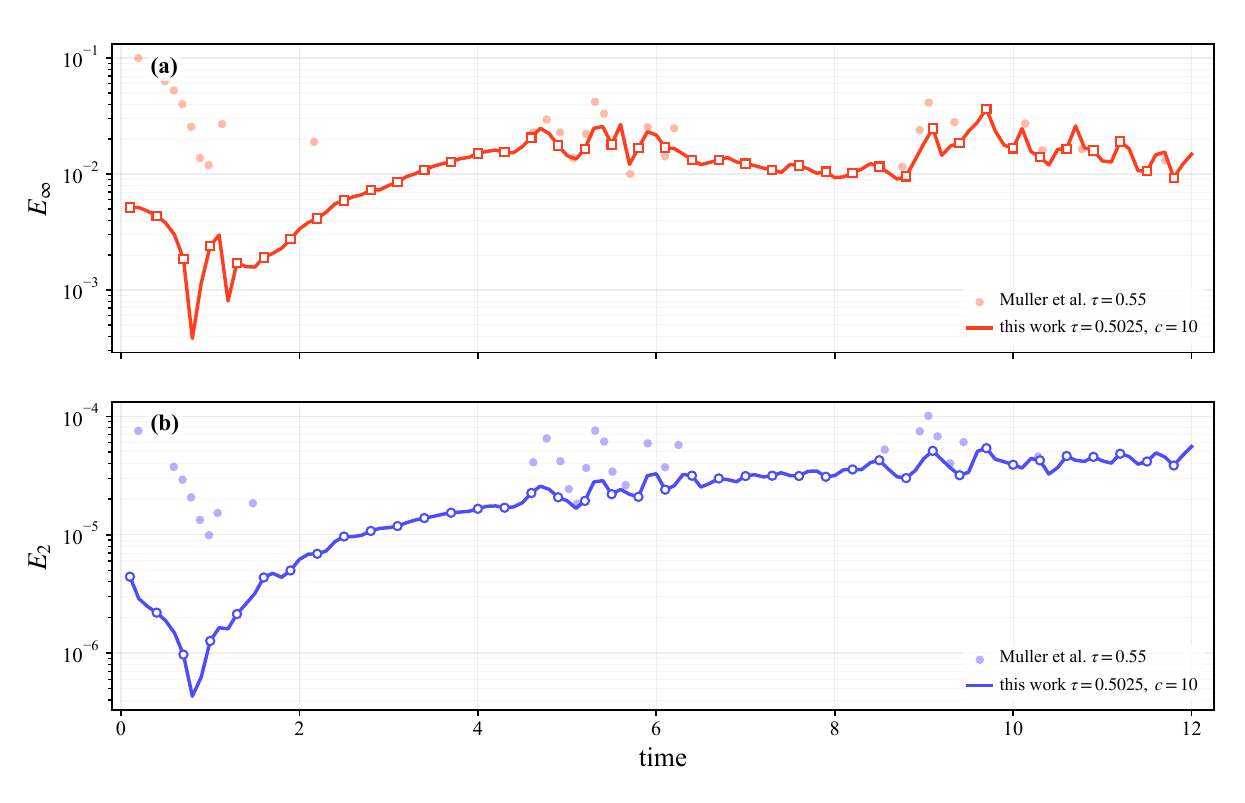}
\caption{Cantilever response compared with a finite-element reference. The present D2Q4\(\times\)6 errors are measured against an independent Q1 reference using the benchmark definitions of \(E_\infty\) and \(E_2\), and are shown together with the \(\tau=0.55\) results of M\"uller et al.~\citep{muller2025nonlinear}.}
\label{fig:wave-beam-error-history}
\end{figure}

Figure~\ref{fig:wave-beam-error-history} reports the time variation of \(E_\infty\) and \(E_2\) up to \(t=12\), together with the \(\tau=0.55\) results of M\"uller et al.~\citep{muller2025nonlinear}. With \(\tau=0.5025\) and \(c=10\), the maximum errors are \(E_2=5.55\times10^{-5}\) and \(E_\infty=3.66\times10^{-2}\). A companion run at \(\tau=0.55\) gives larger maxima, \(E_2=1.44\times10^{-4}\) and \(E_\infty=8.92\times10^{-2}\), showing the sensitivity of this corner-loaded wave problem to the relaxation parameter. Together with the deformed wave fields in Fig.~\ref{fig:cantilever-wave-snapshots}, this comparison supports the use of the same boundary reconstruction for longer-time bending-wave propagation.

\FloatBarrier

\section{Conclusions and outlook}
\label{sec:conclusions-outlook}

This work has introduced a total-Lagrangian vectorial lattice Boltzmann formulation for two-dimensional finite-strain hyperelastic dynamics. The governing equations were written as a conservative first-order system for the material velocity and the full deformation gradient, allowing the lattice to remain fixed in the reference configuration while the first Piola--Kirchhoff stress is evaluated locally from \(\bF\). A D2Q4 lattice with six-component vector populations was then used to match the macroscopic state and the two material-coordinate fluxes. In this way, finite-strain material nonlinearity enters only through nonlinear flux moments, whereas the method retains the local collide--stream structure of LBM. The formulation further includes trapezoidally centered forcing, second-order population initialization, displacement recovery by velocity quadrature, and half-way reconstructions for velocity Dirichlet and nominal-traction Neumann boundaries on grid-aligned domains.

The numerical results support the accuracy and applicability of the construction. Periodic manufactured solutions show second-order convergence for displacement, first Piola stress, and Cauchy stress when the second-order initialization is used. Boundary manufactured solutions confirm the expected behaviour for Dirichlet data and show that mixed traction boundaries mainly introduce localized stress errors, while the interior remains governed by the second-order bulk discretization. The uniaxial-tension and simple-shear benchmarks further demonstrate that the same flux and boundary reconstructions carry over to finite-strain benchmark configurations. The method also reproduces homogeneous affine stress responses for several hyperelastic laws, acoustic-tensor wave speeds about predeformed states, finite-amplitude periodic shear waves, and a bounded-domain cantilever bending wave.

These results indicate that the proposed total-Lagrangian vectorial LBM can represent both nonlinear constitutive response and elastodynamic wave propagation within one explicit local framework. The formal consistency argument in Appendix~\ref{app:second-order-consistency} supports this observation for smooth solutions: under acoustic scaling, \(\omega=2\), second-order initialization, and sufficiently accurate boundary data, the leading macroscopic field satisfies the target first-order hyperelastic system. Compared with the linear vectorial formulation, the essential change is the use of state-dependent Piola fluxes and tangent moduli; the moment-matching mechanism itself remains unchanged.

Several limitations remain and define natural directions for future work. The present study is restricted to two dimensions, uniform Cartesian lattices, and grid-aligned boundaries. A natural next step is to carry the same construction to three dimensions, where the state would contain three velocity components and the full \(3\times3\) deformation gradient. The treatment of boundaries will also need to move beyond the present grid-aligned setting. For practical solid-mechanics applications, the method should be able to handle curved surfaces, cut cells, corners, and mixed displacement--traction data on arbitrary geometries, while preserving as much as possible the locality that makes LBM attractive. Another promising extension is anisotropic hyperelasticity. In a total-Lagrangian formulation, material directions are fixed in the reference configuration and can therefore be passed directly to the local constitutive update, making orthotropic and fiber-reinforced solids a natural target.

\section*{Data availability}

The data underlying this article are available in Zenodo at \href{https://doi.org/10.5281/zenodo.20376790}{https://doi.org/10.5281/zenodo.20376790}.

\clearpage
\appendix
\section*{Appendix}
\numberwithin{equation}{section}
\section{Constitutive laws used in the implementation}
\label{app:constitutive-laws}

Let
\begin{equation}
  \bC=\bF^T\bF,\qquad
  \bE=\frac{\bC-\Id}{2},\qquad
  J=\det\bF,\qquad
  \bH=\bF^{-T}.
  \label{eq:constitutive-kinematic-quantities}
\end{equation}
For the invariant-based Mooney--Rivlin, Yeoh, and Gent closures, define
\begin{equation}
  I_1=\tr(\bC)+1,\qquad
  I_2=J^2+\tr(\bC),\qquad
  \bar I_1=J^{-2/3}I_1,\qquad
  \bar I_2=J^{-4/3}I_2,\qquad
  A=\bar I_1-3,
  \label{eq:modified-invariants}
\end{equation}
and
\begin{equation}
  \bG_1=\frac{\partial\bar I_1}{\partial\bF}
  =J^{-2/3}\left(2\bF-\frac{2}{3}I_1\bH\right),\qquad
  \bG_2=\frac{\partial\bar I_2}{\partial\bF}
  =J^{-4/3}\left[2\bF+\left(2J^2-\frac{4}{3}I_2\right)\bH\right].
  \label{eq:modified-invariant-gradients}
\end{equation}
The strain energies and first Piola stresses are
\begin{align}
  W_{\rm SVK}
  &=\frac{\lambda}{2}[\tr(\bE)]^2+\mu\,\bE:\bE, \notag\\
  \bP_{\rm SVK}
  &=\bF\bigl[\lambda\tr(\bE)\Id+2\mu\bE\bigr],
  \label{eq:svk-model}
  \\
  W_{\rm NH}
  &=\frac{\mu}{2}\bigl(\tr(\bC)-2-2\log J\bigr)
    +\frac{\lambda}{4}\bigl(J^2-1-2\log J\bigr), \notag\\
  \bP_{\rm NH}
  &=\mu(\bF-\bH)+\frac{\lambda}{2}(J^2-1)\bH,
  \label{eq:nh-model}
  \\
  W_{\log{\rm NH}}
  &=\frac{\mu}{2}\bigl(\tr(\bC)-2-2\log J\bigr)
    +\frac{\lambda}{2}(\log J)^2, \notag\\
  \bP_{\log{\rm NH}}
  &=\mu(\bF-\bH)+\lambda\log J\,\bH,
  \label{eq:lognh-model}
  \\
  W_{\rm MR}
  &=c_{10}(\bar I_1-3)+c_{01}(\bar I_2-3)
    +\frac{\lambda}{2}(\log J)^2, \notag\\
  \bP_{\rm MR}
  &=c_{10}\bG_1+c_{01}\bG_2+\lambda\log J\,\bH,
  \label{eq:mr-model}
  \\
  W_{\rm Yeoh}
  &=c_1A+c_2A^2+c_3A^3+\frac{\lambda}{2}(\log J)^2, \notag\\
  \bP_{\rm Yeoh}
  &=(c_1+2c_2A+3c_3A^2)\bG_1+\lambda\log J\,\bH,
  \label{eq:yeoh-model}
  \\
  W_{\rm Gent}
  &=-\frac{\mu J_m}{2}\log\left(1-\frac{A}{J_m}\right)
    +\frac{\lambda}{2}(\log J)^2, \notag\\
  \bP_{\rm Gent}
  &=\frac{\mu}{2(1-A/J_m)}\bG_1+\lambda\log J\,\bH .
  \label{eq:gent-model}
\end{align}
When material parameters are specified through \(\mu\) and Poisson's ratio \(\nu\), the implementation uses \(\lambda=2\mu\nu/(1-2\nu)\). It sets \(c_{10}+c_{01}=\mu/2\) for Mooney--Rivlin and \(c_1=\mu/2\) for Yeoh unless these parameters are supplied explicitly. Admissible states satisfy \(J>0\), with the additional Gent restriction \(A<J_m\).

\section{Formal second-order consistency}
\label{app:second-order-consistency}

This appendix gives the Chapman--Enskog/Taylor expansion behind the second-order statement used in Section~\ref{sec:d2q4-vectorial-lbm}. All lattice moments, fluxes, source terms, and boundary reconstructions are those already defined in the main text. The expansion assumes smooth fields, acoustic scaling with fixed \(c=\Delta x/\Delta t\), the nondissipative choice \(\omega=2\), the initialization in Eq.~\eqref{eq:init-compact}, and half-way boundary data that are consistent to the order stated below.

\subsection{Bulk Chapman--Enskog expansion}
\label{app:bulk-consistency}

Use \(\Delta t\) as the small parameter and write
\begin{equation}
  \bm f_q=\sum_{m\ge0}\Delta t^m\bm f_q^{(m)},\qquad
  \bU^{\rm num}=\sum_{m\ge0}\Delta t^m\bU^{(m)},\qquad
  \bm f_q^{eq}(\bU^{\rm num})=\sum_{m\ge0}\Delta t^m\bm e_q^{(m)} .
  \label{eq:appendix-asymptotic-expansions}
\end{equation}
The equilibrium expansion is
\begin{align}
  \bm e_q^{(0)}
  &=\bm f_q^{eq}(\bU^{(0)}), \notag\\
  \bm e_q^{(1)}
  &=\left(\bm f_q^{eq}\right)'(\bU^{(0)})\bU^{(1)}, \notag\\
  \bm e_q^{(2)}
  &=\left(\bm f_q^{eq}\right)'(\bU^{(0)})\bU^{(2)}
    +\frac12\left(\bm f_q^{eq}\right)''(\bU^{(0)})
      [\bU^{(1)},\bU^{(1)}].
  \label{eq:appendix-equilibrium-expansion}
\end{align}
The Hessian term is present because the stress fluxes are nonlinear. It does not produce an additional macroscopic consistency term: the zeroth moment identity in Eq.~\eqref{eq:equilibrium-moments} implies \(\sum_q\bm e_q^{(m)}=\bU^{(m)}\) order by order.

From the source-shifted state recovery in Eq.~\eqref{eq:moment-U},
\begin{equation}
  \sum_q\bm f_q^{(0)}=\bU^{(0)},\qquad
  \sum_q\bm f_q^{(1)}=\bU^{(1)}-\frac12\bB,\qquad
  \sum_q\bm f_q^{(m)}=\bU^{(m)}\quad(m\ge2).
  \label{eq:appendix-moment-expansion}
\end{equation}
For \(\omega=2\), the explicit source term in Eq.~\eqref{eq:collision} vanishes and the combined collide--stream equation becomes
\begin{equation}
  \bm f_q(\bX+\bm c_q\Delta t,t+\Delta t)
  =2\bm f_q^{eq}(\bU^{\rm num})-\bm f_q(\bX,t).
  \label{eq:appendix-collide-stream-omega2}
\end{equation}
Expanding the left-hand side along the lattice characteristic gives
\begin{equation}
  \bm f_q+\Delta tD_q\bm f_q+\frac{\Delta t^2}{2}D_q^2\bm f_q+O(\Delta t^3)
  =2\bm f_q^{eq}(\bU^{\rm num})-\bm f_q ,
  \label{eq:appendix-streaming-taylor}
\end{equation}
with \(D_q\) defined in Eq.~\eqref{eq:init-compact}. Substitution of Eq.~\eqref{eq:appendix-asymptotic-expansions} into Eq.~\eqref{eq:appendix-streaming-taylor} gives the following balances.

At order \(\Delta t^0\),
\begin{equation}
  \bm f_q^{(0)}=\bm e_q^{(0)}=\bm f_q^{eq}(\bU^{(0)}).
  \label{eq:appendix-leading-equilibrium}
\end{equation}
At order \(\Delta t^1\),
\begin{equation}
  D_q\bm f_q^{(0)}
  =2(\bm e_q^{(1)}-\bm f_q^{(1)}),
  \qquad
  \bm f_q^{(1)}
  =\bm e_q^{(1)}-\frac12D_q\bm e_q^{(0)}.
  \label{eq:appendix-first-order-pop}
\end{equation}
Taking the zeroth moment of Eq.~\eqref{eq:appendix-first-order-pop} gives
\begin{align}
  \sum_qD_q\bm f_q^{(0)}
  &=
  \partial_t\sum_q\bm f_q^{(0)}
  +\partial_X\sum_qci\,\bm f_q^{(0)}
  +\partial_Y\sum_qcj\,\bm f_q^{(0)} \notag\\
  &=
  \partial_t\bU^{(0)}
  +\partial_X\bPhi_X(\bU^{(0)})
  +\partial_Y\bPhi_Y(\bU^{(0)}),
  \label{eq:appendix-leading-left-moment}
\end{align}
where Eq.~\eqref{eq:equilibrium-moments} has been used with Eq.~\eqref{eq:appendix-leading-equilibrium}. The right-hand side is
\begin{equation}
  2\left(\sum_q\bm e_q^{(1)}-\sum_q\bm f_q^{(1)}\right)
  =2\left(\bU^{(1)}-\bU^{(1)}+\frac12\bB\right)
  =\bB.
  \label{eq:appendix-leading-right-moment}
\end{equation}
Thus the leading field satisfies
\begin{equation}
  \partial_t\bU^{(0)}
  +\partial_X\bPhi_X(\bU^{(0)})
  +\partial_Y\bPhi_Y(\bU^{(0)})
  =\bB.
  \label{eq:appendix-leading-continuum}
\end{equation}

At order \(\Delta t^2\),
\begin{equation}
  D_q\bm f_q^{(1)}+\frac12D_q^2\bm f_q^{(0)}
  =2(\bm e_q^{(2)}-\bm f_q^{(2)}).
  \label{eq:appendix-second-order-pop}
\end{equation}
Using Eqs.~\eqref{eq:appendix-leading-equilibrium} and \eqref{eq:appendix-first-order-pop}, the left-hand side reduces exactly:
\begin{equation}
  D_q\bm f_q^{(1)}+\frac12D_q^2\bm f_q^{(0)}
  =
  D_q\left(\bm e_q^{(1)}-\frac12D_q\bm e_q^{(0)}\right)
  +\frac12D_q^2\bm e_q^{(0)}
  =D_q\bm e_q^{(1)}.
  \label{eq:appendix-second-order-cancellation}
\end{equation}
Taking the zeroth moment of Eq.~\eqref{eq:appendix-second-order-pop} therefore gives
\begin{equation}
  \sum_qD_q\bm e_q^{(1)}
  =2\left(\sum_q\bm e_q^{(2)}-\sum_q\bm f_q^{(2)}\right)
  =\bzero.
  \label{eq:appendix-error-moment}
\end{equation}
The differentiated moment identities following from Eq.~\eqref{eq:feq-derivative} give
\begin{equation}
  \sum_q\bm e_q^{(1)}=\bU^{(1)},\qquad
  \sum_qci\,\bm e_q^{(1)}=\bm A_X(\bU^{(0)})\bU^{(1)},\qquad
  \sum_qcj\,\bm e_q^{(1)}=\bm A_Y(\bU^{(0)})\bU^{(1)}.
  \label{eq:appendix-derivative-moments}
\end{equation}
Consequently,
\begin{equation}
  \partial_t\bU^{(1)}
  +\partial_X\!\left(\bm A_X(\bU^{(0)})\bU^{(1)}\right)
  +\partial_Y\!\left(\bm A_Y(\bU^{(0)})\bU^{(1)}\right)
  =\bzero.
  \label{eq:appendix-first-error-equation}
\end{equation}
This is the homogeneous equation for the first correction. The only finite-strain change relative to the linear vectorial calculation is that the Jacobians in Eq.~\eqref{eq:appendix-first-error-equation} are evaluated along \(\bU^{(0)}\); the nonlinear second derivative in Eq.~\eqref{eq:appendix-equilibrium-expansion} cancels from the zeroth-moment equation through the exact equilibrium moment identity.

\subsection{Initialization and half-way boundaries}
\label{app:init-boundary-consistency}

The initialization in Eq.~\eqref{eq:init-compact} removes the first-order initial error. Comparing it with Eqs.~\eqref{eq:appendix-leading-equilibrium}--\eqref{eq:appendix-first-order-pop} at \(t=0\) gives
\begin{equation}
  \left(\bm f_q^{eq}\right)'(\bU_0)\bU^{(1)}(\bX,0)=\bzero .
  \label{eq:appendix-init-error}
\end{equation}
Summation over \(q\), together with the zeroth differentiated moment in Eq.~\eqref{eq:feq-derivative}, yields
\begin{equation}
  \bU^{(1)}(\bX,0)=\bzero.
  \label{eq:appendix-zero-initial-error}
\end{equation}

For a missing direction \(d=-\bN\) at a half-way boundary, Eq.~\eqref{eq:feq} gives the pair identities
\begin{equation}
  \bm f_d^{eq}+\bm f_{-d}^{eq}=\frac12\bU,\qquad
  \bm f_d^{eq}-\bm f_{-d}^{eq}=\frac1c\,d_A\bPhi_A(\bU).
  \label{eq:appendix-pair-identities}
\end{equation}
Thus anti-bounce-back imposes a state component at leading order, whereas bounce-back with a correction imposes the corresponding normal flux component.

For the velocity Dirichlet rule in Eq.~\eqref{eq:dirichlet-boundary-rule}, the anti-bounce-back velocity entries give
\begin{equation}
  \frac12\bv^{(0)}=\frac12\bv_D,\qquad \bv^{(1)}=\bzero
  \label{eq:appendix-dirichlet-velocity-error}
\end{equation}
at the boundary. The bounce-back deformation-gradient entries impose the kinematic normal flux because
\begin{equation}
  \frac1c d_A\Phi_A^{F_{iB}}(\bU^{(0)})
  =\frac{N_Bv_i^{(0)}}{c},
  \label{eq:appendix-dirichlet-kinematic-flux}
\end{equation}
which is exactly the correction used in Eq.~\eqref{eq:dirichlet-boundary-rule}. The first-order part is therefore homogeneous for the error equation.

For the Neumann rule in Eq.~\eqref{eq:neumann-boundary-rule}, the bounce-back velocity entries impose
\begin{equation}
  \frac1c d_A\Phi_A^{v_i}(\bU^{(0)})
  =\frac1c P_{iA}(\bF^{(0)})N_A
  =\frac{\bar T_i}{c}.
  \label{eq:appendix-neumann-traction}
\end{equation}
The anti-bounce-back deformation-gradient entries impose the boundary value \(\bF^b\). Hence this boundary is second-order consistent provided
\begin{equation}
  \bF^b=\bF(\bX_b,t+\Delta t/2)+O(\Delta t^2)
  \label{eq:appendix-boundary-F-accuracy}
\end{equation}
and the local traction inversion defining \(\bF^b\) is solved to the same order. Under this condition the \(O(\Delta t)\) Neumann error data are homogeneous, namely the linearized traction condition \(\delta\bP\,\bN=\bzero\) together with the corresponding state condition on the reconstructed deformation-gradient entries. If the tangential column extrapolation or local Newton solve is only first order, this boundary part no longer supplies the zero first-order data required by the formal second-order boundary argument, although the interior expansion above is unchanged.

With Eq.~\eqref{eq:appendix-zero-initial-error} and homogeneous first-order boundary data, uniqueness of the linearized homogeneous problem in Eq.~\eqref{eq:appendix-first-error-equation} gives \(\bU^{(1)}=\bzero\). The leading field \(\bU^{(0)}\) satisfies the target first-order system by Eq.~\eqref{eq:appendix-leading-continuum}; therefore, for smooth solutions,
\begin{equation}
  \bU^{\rm num}
  =\bU^{(0)}+\Delta t\bU^{(1)}+O(\Delta t^2)
  =\bU+O(\Delta t^2)
  =\bU+O(\Delta x^2).
  \label{eq:appendix-second-order-consistency}
\end{equation}

\end{document}